\documentclass[reprint,amsmath,amssymb,braket,aps,prb,longbibliography]{revtex4-1}
\usepackage{graphicx,color,braket}
\usepackage{dcolumn}
\usepackage{bm}
\usepackage[colorlinks=true,linkcolor=blue]{hyperref}
\usepackage{colortbl}
\begin{document}
	
	\title{Photon drag at the junction between metal and 2d semiconductor}
	
	\author{Dmitry Svintsov}
	\affiliation{Center for Photonics and 2d Materials, Moscow Institute of Physics and Technology, Dolgoprudny 141700, Russia}
	\email{svintcov.da@mipt.ru}
	
	\author{Zhanna Devizorova}
	\affiliation{Center for Photonics and 2d Materials, Moscow Institute of Physics and Technology, Dolgoprudny 141700, Russia}

	\begin{abstract}
		Photon drag represents a mechanism of photocurrent generation wherein the electromagnetic (EM) field momentum is transferred directly to the charge carriers. It is believed to be small by the virtue of low photon momentum compared to the typical momenta of the charge carriers. Here, we show that photon drag becomes particularly strong at the junctions between metals and 2d materials, wherein highly non-uniform local EM fields are generated upon diffraction. To this end, we combine an exact theory of diffraction at 'metal-2d material' junctions with microscopic transport theory of photon drag, and derive the functional dependences of the respective photovoltage on the parameters of EM field and 2d system. The voltage responsivity appears inversely proportional to the electromagnetic frequency $\omega$, the sheet density of charge, and a dimensionless momentum transfer coefficient $\alpha$ which depends only on 2d conductivity in units of light speed $\eta = 2\pi \sigma/c$ and light polarization. For $p$-polarized incident light, the momentum transfer coefficient appears finite even for vanishingly small 2d conductivity $\eta$, which is a consequence of dynamic lightning rod effect. For $s$-polarized incident light, the momentum transfer coefficient scales as $\eta \ln \eta^{-1}$, which stems from long-range dipole radiation of a linear junction. A simple estimate shows that the ratio of thermoelectric and photon drag photovoltages at the junction for $p$-polarization is roughly $\omega\tau_\varepsilon$, where $\tau_\varepsilon$ is the energy relaxation time, while for $s$-polarization the photon drag always dominates over the thermoelectric effect.
	\end{abstract}
	
	\maketitle
	
	\section{Introduction}
	
	Light expresses pressure force on conducting solids. In fully quantum paradigm, this force results from transfer of photon momentum to the charge carriers. In semi-classical description, light pressure results from Lorentz force expressed by the field on electrons. Such momentum transfer can result in measurable photocurrent, known as photon drag (PD)~\cite{Pd_germanium,ganichev1983drag}. Compared to other photovoltaic effects, photon drag has a strong dependence on the direction of incident light. This factor enabled its observation in various two-dimensional semiconductor structures~\cite{Ganichev_drag_graphene,Ganichev_drag_InSb,Ganichev_drag_110} and confirmation of semi-classical light pressure theory.
	
	The photon drag photocurrent is believed to be small due to the negligible magnitude of photon momentum compared to the electronic ones. Under specific interband resonant conditions, the photon drag can be enhanced~\cite{Entin_PD}. Similarly, the drag photocurrent can appear as the unidirectional surface plasmons are propagating along the conductor~\cite{Noginova_plasmon_drag}. Launching of such unidirectional plasmons is achieved with specially designed asymmetric couplers, the simplest example being the metal grating lacking the inversion center above the 2d semiconductor~\cite{Popov_Non_centrosymmetric,Olbrich_Ratchet}. Such photocurrent is often marked as 'plasmon drag'~\cite{Popov_plasmon_drag}, though the underlying electromagnetic forces are the same as in photon drag.
	
	\begin{figure}[ht!]
		\centering
		\includegraphics[width=0.9\linewidth]{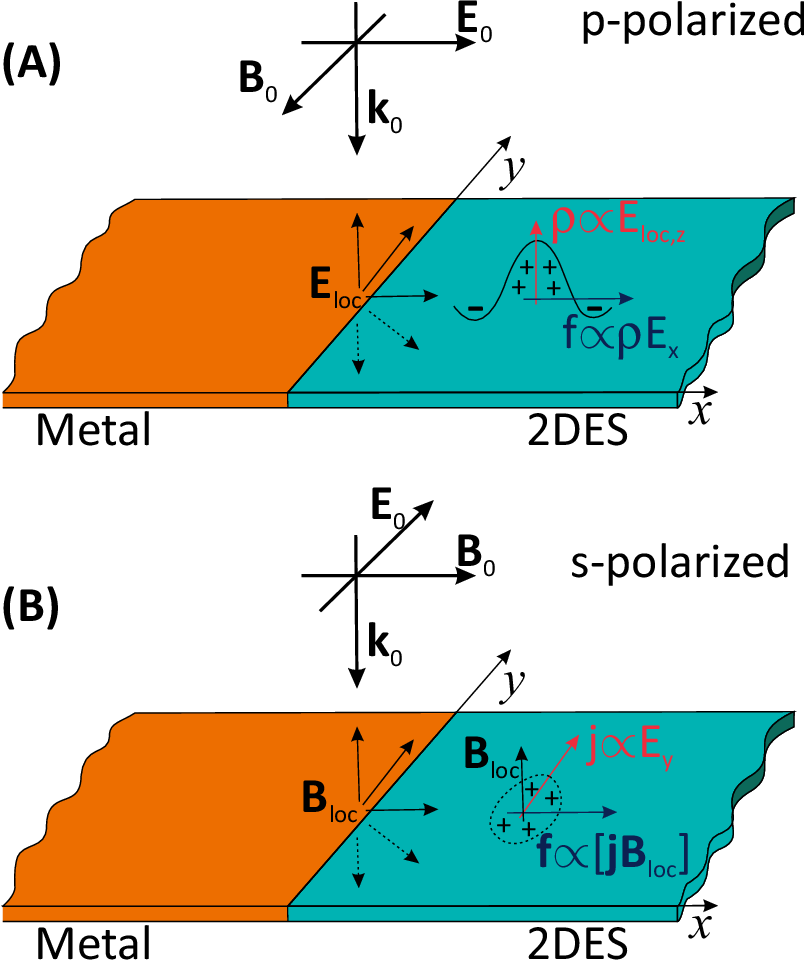}
		\caption{Origin of photon drag at metal-2d semiconductor junction  (a) For $p$-polarized incident wave, scattering at the edge causes local electric fields ${\bf E}_{\rm loc}$ which have both $x$ and $z$ components. The $z$-component induces dynamic charge density $\rho$ in 2DES, while the $x$-component drags the induced charge. (b) For $s$-polarized incident wave, scattering at the edge causes local magnetic fields ${\bf B}_{\rm loc}$ which have both $x$ and $z$ components. $y$-component of electric field induces currents along the junction, which undergo the Lorentz force due to the $z$-component of magnetic field. The Lorentz force is directed outward the junction.}
		\label{fig:structure}
	\end{figure}
	
	Taking a detailed view on Lorentz forces acting on conduction electrons, one can realize that photon drag occurs in generic non-uniform electromagnetic fields. It requires neither co-propagating photons nor well-defined plasmons. Such possibility was addressed in Ref.~\onlinecite{Durnev_structured} which considered the photocurrents induced by light with spatially-structured amplitude and phase. While external structuring of the incident light is possible, another (internal) structuring is ubiquitous and occurs in low-dimensional sub-wavelength optoelectronic devices. Indeed, finite size of conducting channel (e.g. 2d flake or nanotube) and presence of metal contacts implies strong variations of local conductivity. They translate in strong variations of local current density and hence in variations of local fields, as shown in model theoretical studies~\cite{Satou2007,Zabolotnykh_confined} and nicely visualized in scanning near-field imaging~\cite{Titova_Near_Field,Woessner_Nanoscopy,Roskos_SNOM,Alonso-Gonzalez2014,Ni2018}. If the device structure is asymmetric, the average photon drag from these local fields is generally non-zero.
	
	In this paper, we shall show that PD associated with non-uniform local fields is essentially strong at the lateral contacts between metals and two-dimensional semiconductors. Evaluations of the respective photocurrent and photovoltage are important not only as a particular example of enhanced photon drag. Another reasoning for such calculation is ubiquitous presence of metallic contacts in almost all photodetector structures~\cite{Tielrooij_Photocurrent,Tielrooij_Photocurrent_NNano,Ganichev_BLG_subwavelength}. Since Sommerfeld's exact theory of light diffraction by wedges~(\cite{sommerfeld1896}, see also \cite{landau2013electrodynamics,Vainstein,daniele2014wiener} for contemporary approaches), it is known that local electromagnetic field near the bare metal edge is enhanced as $|x|^{-1/2}$, where $x$ is the distance to the edge. We shall show that it is the field singularity that results in elevated photon drag, once the 2d electron system (2DES) is contacted with metal.
	
	The physics of PD at metal-2DES junction depends on light polarization and clarified in Fig.~\ref{fig:structure}. For $p$-polarized incident wave [${\bf B}_0$ is directed along the junction, Fig.~\ref{fig:structure} (A)], electromagnetic diffraction gives rise to the local $z$-component of electric field. This components induces the charge density $\rho(t) \propto E_z$ via the dynamic field effect. The $x$-component of the field drags the induces charges longitudinally, which results in photocurrent. For $p$-polarized incident wave [${\bf E}_0$ is directed along the junction, Fig.~\ref{fig:structure} (B)], electromagnetic diffraction gives rise to the local $z$-component of magnetic field. This component expresses the Lorentz force on conduction currents $j_y$. The force, averaged over fast field oscillations, is non-zero and again results in finite $x$-component of the photocurrent.
	
	The presence of photocurrents due to non-uniform electromagnetic fields in 2d nanostructures was realized in numerous papers under the terms 'resistive self-mixing' and 'Dyakonov-Shur rectification'~\cite{DS_rectification,Sakowicz2011,Roskos_DRSM}. Despite different terminology, these photocurrents represent essentially the photon drag. This becomes clear by writing the expressions for photovoltage in terms of local fields. Unfortunately, all theoretical calculations of respective photovoltage were based either on simplified 'plane wave representations' of the local fields~\cite{DS_rectification,Matyushkin2020,Polini_PW_rectification}, or on electromagnetic simulations~\cite{Fateev_Rectification,ludwig20242d}. These facts obscured the strong photon drag at metal-2DES interface from the view of the researchers.

	\section{Theoretical derivation of photon drag photovoltage}
	Our calculation proceeds in three steps. First, we evaluate the momentum transfer between generically non-uniform electromagnetic field and 2d electron system. Second, we derive the expression for the resulting photovoltage from momentum and energy balance equations. Third, we borrow the results of exact electromagnetic diffraction theory at the junction of metal and 2d semiconductor, namely, the Fourier spectra of the scattered fields~\cite{Nikulin_Edge,svintsov2024exact}, and evaluate the general relation for PD photovoltage by plugging the particular profile of the local field. 
	
	\subsection{Ponderomotive force on 2d electrons}
	The key ingredient of the subsequent derivation is the photon drag force density from EM field ${\bf f}_{\rm pd}$. It equals the time derivative of momentum density received by 2d electrons from the field, ${\bf f}_{\rm pd}= \left( \partial{\bf P}/\partial t \right)_{\rm field}$. This force density can be derived either from electromagnetic field tensor in the presence of conductive medium, or directly by averaging the total Lorentz force density acting on the 2d electrons:
	\begin{equation}
		\label{eq-ponderomotive1}
		{\bf f}_{\rm pd} = {\left( \frac{\partial {\bf P}}{\partial t} \right)}_{\rm field} = \left\langle \rho \left( t \right){\bf E}\left( t \right) \right\rangle +\frac{1}{c}\left\langle \left[ {{\bf j}}\left( t \right),{\bf B}\left( t \right) \right] \right\rangle. 
	\end{equation}
	In the above expression, brakets $\langle...\rangle$ stand for time averaging over fast field oscillations, $\rho(t)$ and ${\bf j}(t)$ are the linear-response charge and current densities in the 2DES which are induced by the field itself. Finally, ${\bf E}\left( t \right)$ and ${\bf B}\left( t \right)$ are the alternating electric and magnetic fields acting on the 2DES which are generally different from the incident field due to the self-consistent screening. The first and second terms in Eq.~(\ref{eq-ponderomotive1}) are relevant to the $p$- and $s$-polarizations of the incident wave, respectively.
	
	We consider the most practical case of monochromatic waves, ${\bf E}(t) = {\bf E}_\omega e^{-i\omega t} + {\rm h.c.}$, and similarly for other time-dependent quantities. The charge and current densities in that case can be expressed with continuity equation and Ohm's law:
	\begin{equation}
		-i\omega \rho_\omega + (\nabla, {\bf j}_{\omega}) = 0, \qquad {\bf j}_{\omega} = \sigma_\omega {\bf E}_{\omega},
	\end{equation} 
	where $\sigma_\omega$ is the dynamic 2DES conductivity. The functional form of $\sigma_\omega$ is not limited to the Drude model at this stage and can be arbitrary. Performing the explicit time averaging in (\ref{eq-ponderomotive1}) we find, after some simplifications:
	\begin{equation}
		\label{eq-ponderomotive2}
		{\bf f}_{\rm pd} =\frac{2}{\omega }\operatorname{Im}{{\sigma }_{\omega }}
		\left\{ \left( { {\nabla^{\parallel}}{{\bf E}}_{\omega }} \right){{\bf E}^{*\parallel}_{\omega }}
		- 
		[ {\bf E}_{\omega }^{\parallel}\times [{\nabla \times}{{\bf E}^{*}_{\omega }} ] ] \right\}.
	\end{equation}
	The superscript $\parallel$ implies taking only the $x$ and $y$ components of the respective vector, it stems from the two-dimensional character of the current density. Further simplifications are possible if all fields do not depend on the coordinate $y$ along the junction. This is eventually true if the incident field has no gliding component of the wave vector, $k_y = 0$. For such polarization state,
	\begin{equation}
		\label{eq-force-final}
		{\bf f}_{\rm pd}
		=
		\frac{2}{\omega }\operatorname{Im}\left\{ {{\sigma }_{\omega }}\left[ \frac{\partial {{E}_{x}}}{\partial x}E_{x}^{*}-\frac{\partial E_{y}^{*}}{\partial x}{E_{y}} \right] \right\},
	\end{equation}
	where we have skipped, for brevity, the Fourier index $\omega$ of all fields. 
	
	Equation (\ref{eq-force-final}) for the force density is applicable for arbitrary variations of fields and dynamic conductivity with coordinate $x$. The only underlying assumption is the locality of current-field relation ${\bf j}_\omega(x) = \sigma_\omega(x) {\bf E}_\omega(x)$, which is valid if the field varies weakly at the electron free path. The convenience of the above expression stems from the fact that the field profiles ${\bf E}(x)$ can be found from linear-response solution of Maxwell equations, e.g. from electromagnetic simulations or exact solutions of the diffraction problems. The evaluation of photocurrent, being the second-order functional in electromagnetic fields, is largely decoupled from the diffraction problem.

	\subsection{Photon drag and thermoelectric voltages}
	
	Known the ponderomotive force density on 2d electrons (\ref{eq-force-final}), we aim to derive the photovoltage generated at the linear junction of metal and 2DES. We start our model with arbitrary spatial dependence of electron density $n(x)$, and pass to a stepwise limit in the final result. The case of metal-contacted 2d electron system would correspond to very large density at $x<0$. 
	
	The local momentum balance equation in the second order with respect to the electromagnetic field reads as
	\begin{equation}
		\label{eq-momentum-balance}
		\frac{\partial {\Pi^{(2)}_{ij}}}{\partial {x_j}}-en\left( x \right)\frac{\partial {{V}_{ph}}}{\partial {x_i}}=
		{\left( \frac{\partial {P_i}}{\partial t} \right)}_{\rm coll}+
		{\left( \frac{\partial {P_i}}{\partial t} \right)}_{\rm field},
	\end{equation}
	where $\Pi^{(2)}_{ij}$ is the second-order field-induced correction to the stress tensor of 2d electrons, $V_{ph}(x)$ is the local photovoltage, and the two terms on the right-hand side describe the momentum loss due to the electron collisions and the momentum gain due to the photon drag, respectively.
	
	Though Eq.~(\ref{eq-momentum-balance}) has no limits of applicability, it can become useful for the evaluation of photocurrent if only the functional forms of momentum vector ${P}_i$ and stress tensor $\Pi^{(2)}_{ij}$ are specified. At this stage, we assume that all angular harmonics of the second-order dc distribution function, except for the zeroth and the first ones, are damped. This occurs either for strong electron-electron collisions~\cite{Ratchet_signature}, or for dominant low-angle Coulomb scattering by charged impurities. In such approximation, $\Pi^{(2)}_{\alpha \beta}$ reduces to a scalar pressure, while its gradient is due to the spatial variations of electron temperature ($C$ is the heat capacitance):
	\begin{equation}
		\label{eq-stress}
		\frac{\partial {\Pi^{(2)}_{ij}}}{\partial {{x}_{\beta }}} = \frac{C}{2} \frac{\partial T}{\partial x_i}.
	\end{equation}
	The momentum loss due to collisions, in the same approximation, is proportional to the drift velocity ${\bf u}$ and inversely proportional to the momentum relaxation time
	\begin{equation}
		\label{eq-collision-loss}
		{\left( \frac{\partial {P_i}}{\partial t} \right)}_{\rm coll} = - \frac{n(x) m u_i (x)}{\tau_p}.
	\end{equation}
	To close the system for determination of photocurrent and photovoltage, we add the thermal balance equation for the electronic subsystem:
	\begin{equation}
		\label{eq-thermal-conduction}
		-\frac{\partial}{\partial x_i} \left( \chi \frac{\partial T}{\partial  x_i}\right) = -C \frac{ T - T_0}{\tau_\varepsilon} + 2 {\rm Re} \sigma_\omega |{\bf E}_\omega|^2.
	\end{equation}
	In the most experimentally relevant situation when all quantities depend only on the $x$-coordinate normal to the junction, and the circuit is open (no photocurrent), the total photovoltage developed across the junction has the form:
	\begin{gather}
		{{V}_{\rm ph}}={{V}_{\rm pte}}+{V_{\rm pd}}, \\ 
		{{V}_{\rm pte}}=-\int\limits_{-\infty }^{+\infty }{S\left( x \right)\frac{dT}{dx}dx}; \\ 
		\label{eq-pd-extended}
		{{V}_{\rm pd}}=-\frac{2}{\omega }\int\limits_{-\infty }^{+\infty }{\frac{1}{e n(x)}\operatorname{Im}\left\{ {{\sigma }_{\omega }}\left[ \frac{\partial {{E}_{x}}}{\partial x}E_{x}^{*}-\frac{\partial E_{y}^{*}}{\partial x}{{E}_{y}} \right] \right\}dx}.	
	\end{gather}
	Here, $V_{\rm pte}$ and $V_{\rm pd}$ are the partial photovoltages due to the thermoelectric effect and photon drag, respectively, $S(x)$ is the position-dependent Seebeck coefficient, and the integration is performed from the leftmost contact to the rightmost one, which, within the theory, can be moved to infinity. We consider now the limit of metallic contact to the 2DES, i.e. $n(x<0) \rightarrow \infty$. Demoting the Seebeck coefficient and carrier density in 2DES as $S$ and $n_{\rm 2d}$, we find
	\begin{gather}
		\label{eq-pte}
		V_{\rm pte} = S [T(x=0) - T_0],\\
		\label{eq-pd}
		{V_{\rm pd}}=-\frac{2}{\omega e n_{\rm 2d}} \int\limits_{0}^{+\infty } {\operatorname{Im}\left\{ {{\sigma }_{\omega }}\left[ \frac{\partial {{E}_{x}}}{\partial x}E_{x}^{*}-\frac{\partial E_{y}^{*}}{\partial x}{{E}_{y}} \right] \right\}dx}.	
	\end{gather}
	
	The thermoelectric voltage (\ref{eq-pte}) gained considerable attention in the literature on photoresponse in graphene contacts~\cite{Herrero_TE,Levitov_PTE,Tielrooij_Photocurrent}, while the contribution of photon drag was often overlooked therein. On the contrary, the literature on photoresponse in terahertz field-effect transistors~\cite{DS_rectification,Polini_PW_rectification,Fateev_Rectification}, with rare exceptions~\cite{Ivchenko_heating_in_FETs}, ignored the contribution of thermoelectric effect. Our result (\ref{eq-pte}, \ref{eq-pd}) coincides with the previously derived photovoltage generated in 2DES by structured light \cite{Durnev_structured} with two caveats. First, our consideration excludes the photocurrent due to the excitation of second angular harmonic of distribution function by light. This contribution can be readily restored and considered in line with photon drag for our particular geometry. Second, it is essential to apply the imaginary part to the product $\sigma_\omega E^* \partial_x E$ in our case of confined 2DES. When the 2DES is unbounded and the light beam is confined, the photon drag photovoltage becomes proportional to ${\rm Re}\sigma_\omega$, as in Ref.~\cite{Durnev_structured}~\footnote{To prove this, we denote $I = \int_{-\infty}^{+\infty}{E^* \partial_x E dx}$. The expression for photon drag voltage across the extended 2DES is proportional to ${\rm Im} (\sigma_\omega I) = {\rm Im} \sigma_\omega {\rm Re} I + {\rm Re} \sigma_\omega {\rm Im} I$. To show that the first term is zero, we write ${\rm Re} I = (\int_{-\infty}^{+\infty}{E^* \partial_x E dx} + \int_{-\infty}^{+\infty}{E \partial_x E^* dx})/2$. Integrating the second summand by parts, we get ${\rm Re} I = |E|^2|_{-\infty}^{+\infty}/2$ which is zero for confined light beam and extended 2DES. Apparently, this trick does not apply to confined 2DES with non-zero field at the periphery.}.

	\subsection{Photon drag by local fields at 'graphene - 2DES' junctions}
	
	Evaluation of photon drag photovoltage at 'metal - 2DES' junctions requires the knowledge of local electric fields ${\bf E}(x)$. The explicit form of their Fourier spectra ${\bf E}(q)$ was found in Refs.~\cite{Nikulin_Edge,svintsov2024exact} using the Wiener-Hopf technique for electromagnetic scattering problem. The results depend on the polarization of the incident wave. For $p$-polarization (also referred to as TM, Fig.~\ref{fig:structure} A), where the incident magnetic field ${\bf H}_0$ is parallel to the junction, the local electric field has the form:
	\begin{equation}
		\label{eq-electric-p}
		E_{x}(q)= -\frac{i E_0 \sin \theta }{(q-k_x - i \epsilon)(1+\eta \sin{\theta})} \frac{\varepsilon^{\rm p}_{-}(k_x) \sqrt{k_0 - q}}{\varepsilon^{\rm p}_{-}(q) \sqrt{k_0 - k_x}},
	\end{equation}
	and the $y$-component is absent. Above, $\varepsilon^{\rm p}(q)$ is the dielectric function of 2DES for $p$-polarization,
	\begin{equation}
		\varepsilon^{\rm p}(q)=1 + \eta \frac { \sqrt { k_0^{ 2 } - q^{2}}} { k_0 }  , \qquad \eta = \frac{2\pi\sigma}{c},
	\end{equation}
	and $\varepsilon^{\rm p}_-(q)$ is the part of 2DES dielectric function which is analytic in the lower half-plane of complex $q$-variable (see Appendix A for explicit form). In all above expressions, $k_0$ is assumed to have small imaginary part. 
	
	For $s$-polarization (also referred to as TE, Fig.~\ref{fig:structure} B), where the incident electric field ${\bf E}_0$ is parallel to the junction, the local field has the form:
	\begin{equation}
		\label{eq-electric-s}
		E_y \left( q \right)=\frac{-i{{E}_{0}}}{(1+{{\eta }}/\sin \theta)(q- {k}_{x} - i\epsilon  ) }\frac{\sqrt{{{k}_{0}}-{{k}_{x}}}}{\sqrt{k_0-q}}\frac{\varepsilon^{\rm s}  _{-}\left( {{k}_{x}} \right)}{\varepsilon^{\rm s}_{-}\left( q \right)},
	\end{equation}
	where $\varepsilon^{\rm s}(q)$ is the dielectric function of 2DES for $s$-polarization
	\begin{equation}
		\varepsilon^{\rm s}(q)=1 + \eta \frac { k_0} { \sqrt { k_0^{ 2 } - q^{2}} },
	\end{equation}
	and $\varepsilon^{\rm s}_-(q)$, as before, is obtained by factorization of dielectric function in the complex $q$-variable.
	
	Equation (\ref{eq-pd}) for photon drag photovoltage supplemented by expressions of the local electric fields (\ref{eq-electric-p},\ref{eq-electric-s}) form the necessary building blocks for numerical evaluations. The results can be cast in a transparent form by introducing the photovoltage responsivity of the junction per incident light intensity $r_{\rm pd} = V_{\rm pd} Z_0 /(2 E_0^2)$, where $Z_0 = 4\pi/c$ is the free-space impedance. This results in
	\begin{gather}
		\label{eq-rv-drag}
		r_{\rm pd} = - \frac{2}{\omega e n_{\rm 2d}} \left(\alpha_x + \alpha_y \right),\\
		\label{eq-rv-integral}
		\alpha_i = \frac{1}{E_0^2}{\rm Im}\left\{\eta \int\limits_{0}^{\infty}{\frac{\partial {E_i}}{\partial x}E_{i}^{*} dx}\right\},\qquad i=\{x,y\}.
	\end{gather}
	The quantities $\alpha_i$ can be called 'momentum transfer coefficients', in analogy with absorption coefficients in the problem of energy transfer. They are dimensionless and depend only on 2d conductivity $\eta$ and angle of incidence $\theta$. It is ultimately convenient to express $\alpha_i$ via the Fourier coefficients of the field:
	\begin{equation}
		{{\alpha }_{i}}=\frac{1}{E_{0}^{2}}\left\{ \int\limits_{-\infty }^{+\infty }{iq{{\left| {{E}_{i}}\left( q \right) \right|}^{2}}\frac{dq}{2\pi }}-\frac{1}{2}{{\left| {{E}_{i}}\left( x=0 \right) \right|}^{2}} \right\},
	\end{equation}
	the term $E_i(x=0)$ comes from discontinuity of electric field at metal-2DES interface and is relevant only to the $p$-polarization. In further analysis, we will study the functional dependences of momentum transfer coefficients on 2D conductivity. As for the dimensional prefactor $2/(\omega e n_{\rm 2d})$ in the expression for responsivity, its typical values vary between 20 $\mu$V cm$^2$/W at $n_{\rm 2d} = 10^{11}$ cm$^{-2}$ and $\omega/2\pi = 1$ THz and 0.7 $\mu$V cm$^2$/W at $\omega/2\pi = 30$ THz ($\lambda = 10$ $\mu$m).

	\section{Results: polarization and conductivity-dependent photon drag}
	To highlight the importance of photon drag induced by local fields, we consider the case of normal incidence $k_x = 0$. The contribution of the ordinary drag due to finiteness of $k_x$ will be estimated in the discussion.
	
\begin{figure}[ht!]
		\centering
		\includegraphics[width=0.85\linewidth]{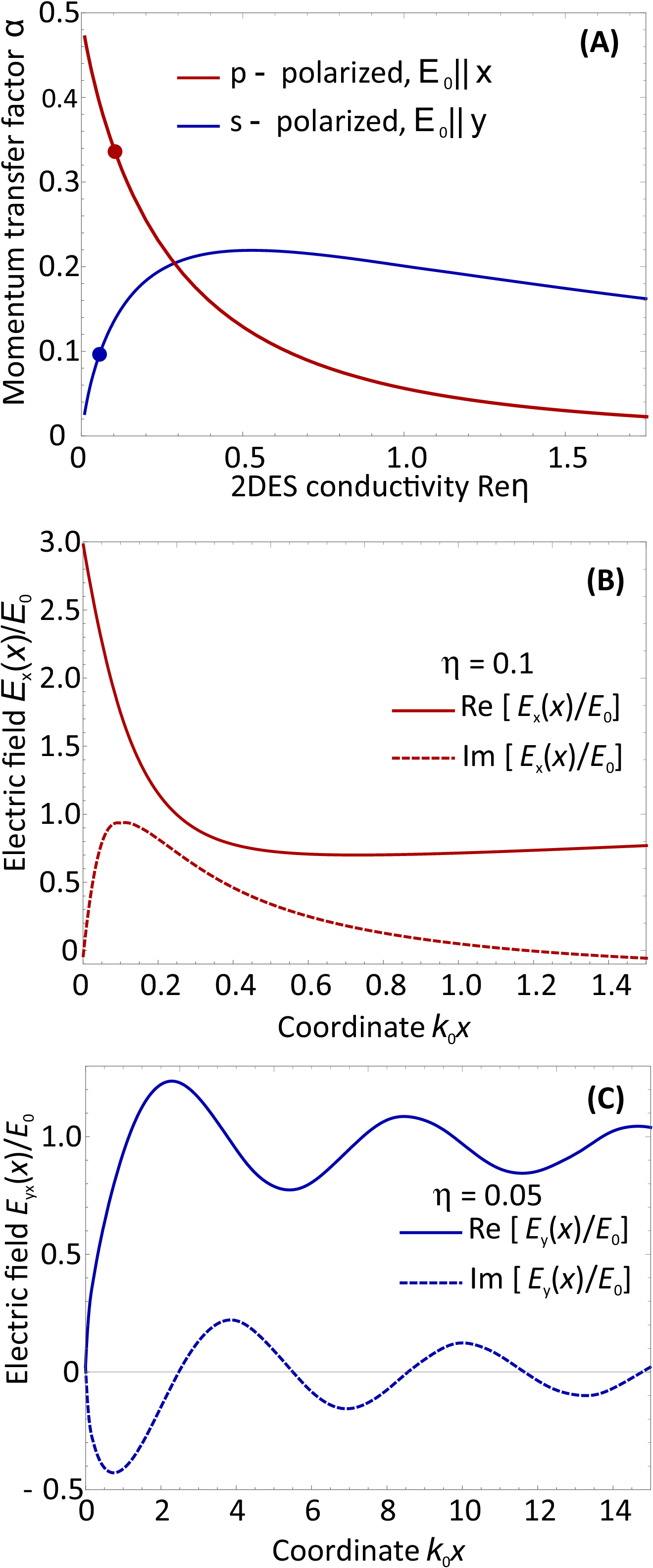}
		\caption{Photon drag at metal-2d semiconductor interface for real 2DES conductivity. (A) Dependence of momentum transfer coefficient on 2DES conductivity (assumed real) for two polarizations of incident light. With reducing $\eta$, the coefficient for $p$-polarized wave tends to a constant. This is explained by singular growth of electric field at the interface with reducing conductivity. An example of field profile is shown in (B) for $\eta = 0.1$. For $s$-polarized wave, the momentum transfer scales as $\eta \ln \eta^{-1}$. The log-divergence is associated with slowly decaying cylindrical wave $E_y \propto e^{ikx}/x^{1/2}$ emitted by metallic half-plane in the absence of 2DES. An example of the field profile is shown in (C) for $\eta = 0.05$}
		\label{fig:pd_real}
	\end{figure}

	We start our numerical analysis of momentum transfer coefficients for the real conductivity of the 2DES. The case is relevant to graphene at infrared frequencies with 'unlocked' interband transitions, $\eta \approx 1.15 \times 10^{-2}$. The functional dependences of $\alpha_i$ on real $\eta$ are shown in Fig.~\ref{fig:pd_real} (A). Both transfer coefficients go to zero with increasing $\eta$, which results from the trivial 'mirror' effect by highly conductive surfaces. More remarkable is the behavior of $\alpha_{i}$ with reducing $\eta$. The coefficient for $p$-polarized wave tends to a constant $\alpha_x \approx 0.45$ at $\eta \rightarrow 0$. The coefficient for $s$-polarized wave demonstrates a superliner behavior $\alpha_y \propto \eta \ln \eta^{-1}$ at small $\eta$. These scalings are highly non-trivial. Indeed, the absence of 2DES implies the absence of momentum transfer, and a linear decay of transfer coefficients at small $\eta$ could be expected.
	
	Insights into these relatively large coefficients of momentum transfer are obtained by inspecting the real-space profiles of the diffracted field. These are shown in Fig.~\ref{fig:pd_real} (B) and (C) for $p$- and $s$-polarizations, respectively, the corresponding values of 2DES conductivity are marked in (A) with circles. As the 2DES conductivity goes to zero, the $p$-polarized field tends to divergence at $x=0$ following
	\begin{equation}
		\label{eq-field-enhancement}
		E_x(x=0) = \frac{E_0}{\sqrt{\eta(1+\eta)}}.
	\end{equation}
	This divergence is the consequence of polarization-dependent lightning-rod effect. In the absence of 2DES, the field near the metal edge would diverge as $x^{-1/2}$~\cite{landau2013electrodynamics}. For small 2DES conductivity, this divergence is smeared according to (\ref{eq-field-enhancement}). The behavior of divergence is such that the product of conductivity $\eta$ and a quadratic function of field remains finite. It is instructive that the area of field enhancement for $p$-polarization spans at a very short length $l \sim \eta \lambda_0$. We can anticipate that the presence of 2DES inhomogenities, other contacts, etc at larger distances would not affect the magnitude of $\alpha_x$.
	
	The momentum transfer factor for $s$-polarization goes to zero at very small $\eta$, which is a more anticipated behavior, in agreement with the absence of field singularities. A more careful examination shows that momentum transfer has a diverging derivative at $\eta \rightarrow 0$, and scales as $\alpha_s \propto \eta \ln \eta^{-1}$. The origin of this super-linear behavior can be traced to the structure of the scattered field at large distances. An example of such structure is shown in Fig.~\ref{fig:pd_real} (C) for $\eta = 0.05$. The large-distance asymptotes of the field are well described by $E_y \propto e^{ik_0x}/x^{1/2}$, which corresponds to a cylindrical wave emitted by the metallic half-plane. This scaling, in the absence of 2DES, fully agrees with prior exact theories of diffraction at metal half-plane~\cite{sommerfeld1896,Senior}. The integral (\ref{eq-rv-integral}) evaluated on this cylindrical wave field diverges logarithmically at large distances. This divergence is cut off due to finite absorbance in 2DES at distances $L \sim \lambda_0 / \eta$, which yields the necessary scaling. In contrast with the case of $p$-polarization, the momentum transfer in $s$-polarization occurs mainly at large distances. Therefore, its clear observation is possible only in structures with length much exceeding $\lambda_0$.

	\begin{figure}[ht!]
		\centering
		\includegraphics[width=0.85\linewidth]{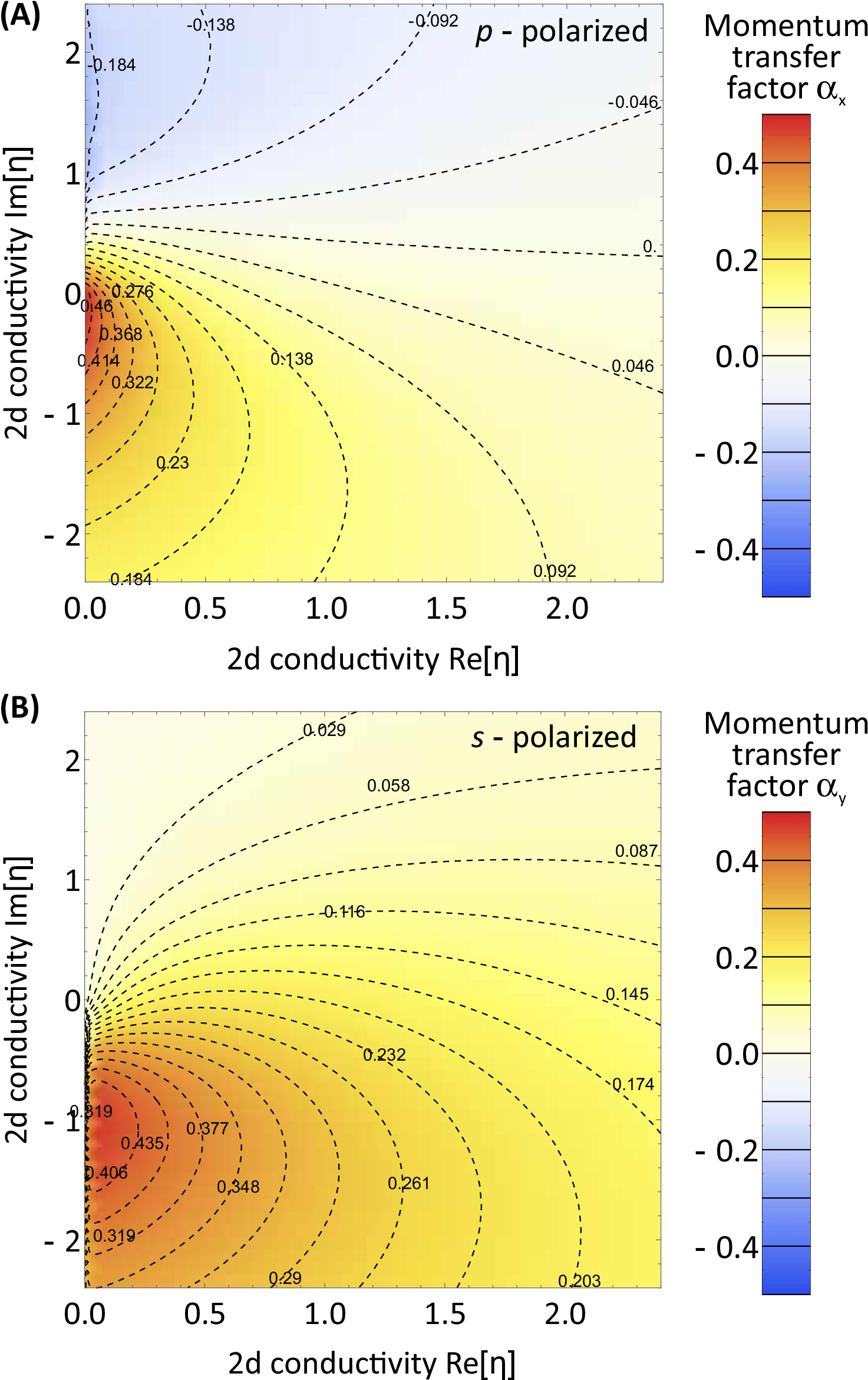}
		\caption{Momentum transfer coefficients at metal-2DES junction $\alpha_x$ (A) and $\alpha_y$ (B) vs real and imaginary parts of 2DES conductivity. Values of $\alpha_{x,y}$ are shown with color and indicated with numbers across the level lines}
		\label{fig:maps}
	\end{figure}
	
	We further inspect the momentum transfer coefficients for complex 2d conductivity $\eta = \eta' + i \eta''$. The results for two polarizations are shown in Fig.~\ref{fig:maps} as color maps. There are two principal differences between momentum transfer in $s$- and $p$-polarizations. First, $\alpha_y$ is always positive, i.e. the drag force is directed outward from the metal for $s$-polarization, while $\alpha_x$ can change sign. Second, the drag force is vanishing for dissipationless 2DES for $s$-polarization, and remains finite for $p$-polarization. Zero drag force for $\eta'=0$ in $s$-polarization is related to zero electric fields at the interface $x=0$.
	
	Large imaginary part of 2DES conductivity stimulates the launching of propagating plasmons away from the edge, as visualized in numerous experiments on near-field microscopy~\cite{Alonso-Gonzalez2014,Ni2018}. These plasmonic oscillations are explicitly seen in the coordinate-dependent field profiles~\cite{Nikulin_Edge}. In the Fourier space, they can be seen as singularities of ${\bf E}(q)$ appearing provided $\varepsilon(q) = 0$. The difference in effective dielectric functions in $s$- and $p$-polarizations results in different plasmon dispersion laws. For $p$-polarization the plasmons exist for inductive 2DES conductivity and possess wave vector $q^{\rm p}_{\rm pl} = k_0[1-\eta^{-2}]^{1/2}$.  For $s$-polarization, the plasmons exist for capacitive 2DES conductivity and possess wave vector $q^{\rm s}_{\rm pl} = k_0[1+\eta^{2}]^{1/2}$~\cite{Mikhailov_new_mode}.
	
	It is remarkable that the existence of 2d plasmons and even their high quality factor does not result in any pronounced enhancement of plasmon drag. The coefficients $\alpha_{x,y}$ remain finite and non-singular across the whole plane of complex 2DES conductivities. This can be explicitly shown for $p$-polarization by substituting the trial plasmonic field profile
	\begin{equation}
		\label{eq-plasmonic-trial}
		E_x = E_{\rm pl} \exp\left( i q^{\rm p}_{\rm pl} x \right) 
	\end{equation}
	into Eq.~(\ref{eq-rv-integral}) and evaluating the respective momentum transfer coefficient. This results in 
	\begin{equation}
		\alpha_x = \frac{|E_{\rm pl}|^2}{|E_0|^2} \left\{ {\rm Re}\eta \times Q_{\rm pl} - \frac{{\rm Im}\eta}{2}\right\},
	\end{equation}
	where $ Q_{\rm pl} = {\rm Re} q^{\rm p}_{\rm pl}/{\rm Im} q^{\rm p}_{\rm pl}$ is the quality factor of 2D plasmons. The above simple calculation explains that plasmonic $Q$-factor (benefiting from low dissipation) appears multiplied by the dissipative part of 2d conductivity ${\rm Re}\eta$. Therefore, no plasmonic enhancement is expected for the drag photovoltage under these resonant conditions. The underlying reasons are twofold. First, the plasma wave should be absorbed to be converted into photovoltage, which requires finite dissipative conductivity. Second, the setup with a single metal contact and long 2DES is not confined in the $x$-direction. As a result, the Fabry-Perot interference which led to plasmonic enhancement of electric field at selected frequencies~\cite{DS_rectification}, is absent in this setup.
	
	Such simple analysis of 'plasmonic drag' is impossible for $s$-polarization, where the trial field (\ref{eq-plasmonic-trial}) does not satisfy the boundary condition of zero field at the interface. A maximum in the drag coefficient $\alpha_y$ is seen at ${\rm Re}\eta \ll 1$ and ${\rm Im}\eta \sim -1$, which corresponds to the launching of long-range TE plasmons. However, this maximum is smooth and not proportional to the TE plasmon quality factor.
	
	\section{Discussion and conclusion}
	
	It is reasonable to compare photon drag photovoltage at the metal-2DES interface with photoelectric effects of different origin. For a long time, the photo-thermoelectric effect at such junctions was considered as dominant~\cite{Tielrooij_Photocurrent,Levitov_PTE}. We estimate is magnitude using Eq.~(\ref{eq-pte}) for photovoltage and local approximation to Eq.~(\ref{eq-thermal-conduction}) for electron temperature. This results in
	\begin{equation}
		r_{\rm pte} = \frac{2\tau_\varepsilon}{en_{2d}} \eta' \left| \frac{E(x=0)}{E_0}\right|^2.
	\end{equation}
	
	Expectedly, the thermoelectric effect is proportional to the local junction field. Therefore, it is small for $s$-polarization and large for $p$-polarization. Using (\ref{eq-field-enhancement}) for local field, and assuming real conductivity, we find
	\begin{equation}
		r_{\rm pte} = \frac{2\tau_\varepsilon}{en_{2d}} = \frac{\omega \tau_\varepsilon}{\alpha_x} r_{\rm pd}.
	\end{equation}
	As the momentum transfer coefficient $\alpha_x \sim 1$, we observe that the only parameter governing the ratio of PTE and PD photovoltages is $\omega\tau_\varepsilon$. The energy relaxation time $\tau_\varepsilon$ varies between hundreds of femtoseconds at room temperature to tens of picoseconds at liquid helium temperature~\cite{Low2012}. Therefore, the junction photon drag can readily exceed the thermoelectric effect, at least at terahertz frequencies. In the far infrared range ($\lambda_0 = 10$ $\mu$m), we estimate $\omega\tau_\varepsilon \gg 1$ even for the shortest energy relaxation time $\tau_\varepsilon = 0.1$ ps.
	
	More accurate estimates of thermoelectric and photon drag photovoltages at metal-2DES junction require the following consideration. The metal contact dopes 2DES by the virtue of work function difference. The carrier density and other kinetic coefficients of 2DES become spatially-dependent in this near-junction layer~\cite{Xia2009a}. In the simplest approximation, the metal-induced carrier density can be considered as constant in the screening layer of length $l_j \sim 100$ nm. Relative polarity of carriers in the screening layer and in the 2DES bulk affects the photoelectric effects. Dissimilar doping ($p-n$ or $n-p$ junction between screening layer and bulk) enhances the PTE and reduces PD, while similar doping ($p-p$ or $n-n$) reduces the PTE and enhances the PD~\cite{Bandurin2018}. Evaluation of photovoltages in this realistic situation can be based on general expressions (\ref{eq-pd}) and (\ref{eq-pte}) combined with simulations of local electric fields.
	
	All our consideration was based on assumption that the photovoltage at an individual metal contact (say, source) can be isolated from the photovoltage at the drain. It means that the incident radiation should be focused to the spot of width $w$ smaller than source-drain separation, $w \ll L_{\rm sd}$. Such illumination is often used experimentally for photocurrent mapping~\cite{Semkin_APL,Xia2009a,Levitov_PTE}. The solution for local field (\ref{eq-electric-p}, \ref{eq-electric-s}) should be modified for confined beams by convolving with the spatial spectrum $s(k_x)$ of  the incident light. Assuming a rectangular aperture of width $w \gg \lambda_0$, this convolution can be performed analytically and results in the replacement
	\begin{equation}
		\frac{1}{q-k_x - i\epsilon} \rightarrow \frac{1-e^{-i(q-k_x)w}}{q-k_x}
	\end{equation}
	in Eqs.~(\ref{eq-electric-p}) and (\ref{eq-electric-s}). Such replacement is generally unimportant for the considered problem of near-field photon drag, as the relevant Fourier components lie at $q\sim k_0 \gg w^{-1}$ and $q\sim q_{\rm pl} \gg w^{-1}$. It can become important only for obtaining a convergent result for the 'ordinary' photon drag emerging in the bulk upon oblique illumination.
	
	Finally, we estimate the 'ordinary' photon drag photoresponsivity $r^{(0)}_{\rm pd}$ upon the tilted illumination of 2DES with finite $k_x$ and finite length $L$. Applying (\ref{eq-pd-extended}) to the field ${\bf E} \propto e^{ik_x x}$, we find
	\begin{equation}
		r^{(0)}_{\rm pd} = -\frac{2}{\omega e n_{2d}} \eta' (k_x L) \frac{|E_x|^2 + |E_y|^2}{|E_0|^2} \sim \eta' (k_x L) r_{\rm pd}.
	\end{equation}
	We observe that 'ordinary' photon drag differs from the 'junction drag' by factors of $\eta'$ (absorbance factor) and $k_x L_{\rm sd}$ (light momentum factor). The former is well below unity for graphene and most quantum wells. The latter is below unity for sub-wavelength devices. An exclusive case where the bulk photon drag can dominate over the junction drag is represented by large-area GaAs quantum wells with 'relativistic' conductivity $\eta' \sim 1$~\cite{Gusikhin2018}.
	
	In conclusion, we have developed a theory of photon drag at an interface between metal and two-dimensional semiconductor. The drag appears due to the local structuring of the near electromagnetic field upon scattering at the metal contact. Consideration of these field modifications and this sort of drag should be important upon interpretation of terahertz detection experiments with sub-wavelength structures~\cite{Ganichev_BLG_subwavelength}. The resulting 'contact drag photovoltage' does not contain typical smallness parameters associated with low photon momentum and low absorbance by (most) two-dimensional systems.  We attempted to formulate the theory in terms of high-frequency 2d conductivity $\sigma_\omega$, independent of its microscopic nature. Thus, the theory should be applicable to various 2DES, including graphene, III-V and II-VI quantum wells, and transition metal chalcogenides.
	
	\section{Acknowledgement}
	The work was supported by the grant \# 21-79-20225 of the Russian Science Foundation. D.S. is grateful to Mikhail V. Durnev for pointing to the possibility of photon drag at the metal-2DES interfaces.
	
	\appendix
	\section{Factorized dielectric functions}
	Factorization of the dielectric function $\varepsilon(q)$ is achieved by applying the Cauchy theorem of complex analysis to a narrow strip of width $2\gamma$ enclosing the real $q$-axis. This result is
	\begin{equation}
		\label{eq-cauchy-factorization}
		\varepsilon _{\pm } \left( q \right)=\exp \left\{ \pm \frac{1}{2\pi i}\int\limits_{-\infty }^{+\infty }{\frac{\ln \varepsilon \left( u \right)du}{u-\left( q\pm i\gamma  \right)}} \right\}.
	\end{equation}
	It applies to both dielectric functions in the $s$- and $p$-polarizations. It is possible to relate the factorized dielectric functions in the two polarizations using an algebraic identity
	\begin{equation}
		\label{eq-s-p-relation}
		1 + \eta \frac { \sqrt { k_0^{ 2 } - q^{2}}} { k_0 }  = \frac{\eta k_0}{\sqrt { k_0^{ 2 } - q^{2}}} \left(1 + \frac{1}{\eta} \frac { \sqrt { k_0^{ 2 } - q^{2}}} { k_0 }\right).
	\end{equation}
	Applying the factorization procedure to both sides, we get
	\begin{equation}
		\varepsilon^{\rm s}_\pm(q,\eta) = \frac{\sqrt{\eta k_0}}{\sqrt{k_0 \pm q}} \varepsilon^{\rm p}_\pm(q,\eta^{-1}).
	\end{equation}
	It implies that only one non-trivial factorization is required, say, for $\varepsilon^{\rm p}_\pm(q)$. Analytical evaluation of $\varepsilon^{\rm p}_\pm(q)$ via dilogarithm functions is possible and is presented in~\cite{Nikulin_Edge}. Alternatively,  a direct numerical evaluation using Eq.~(\ref{eq-cauchy-factorization}) is possible. When dealing with $q$-values on the real axis (which is sufficient for all calculations), we rewrite Eq.~\ref{eq-cauchy-factorization} as:
	\begin{multline}
		\label{eq-factorization-num}
		\varepsilon _{\pm } \left( q \right)=\sqrt{\varepsilon(q)} \times \\
		\exp \left\{ \pm \int\limits_{0}^{+\infty }{\frac{\ln \varepsilon \left( q +v \right) - \ln \varepsilon \left( q - v \right)}{ 2\pi i v} dv} \right\}.
	\end{multline}
	The convenience of representation (\ref{eq-factorization-num}) stems from removal of integration singularities at zero and infinity. The integral converges in an ordinary sense (not only as principal value), and simple grid integration methods can be applied.

	\bibliography{sample}

\begin{thebibliography}{45}%
\makeatletter
\providecommand \@ifxundefined [1]{%
 \@ifx{#1\undefined}
}%
\providecommand \@ifnum [1]{%
 \ifnum #1\expandafter \@firstoftwo
 \else \expandafter \@secondoftwo
 \fi
}%
\providecommand \@ifx [1]{%
 \ifx #1\expandafter \@firstoftwo
 \else \expandafter \@secondoftwo
 \fi
}%
\providecommand \natexlab [1]{#1}%
\providecommand \enquote  [1]{``#1''}%
\providecommand \bibnamefont  [1]{#1}%
\providecommand \bibfnamefont [1]{#1}%
\providecommand \citenamefont [1]{#1}%
\providecommand \href@noop [0]{\@secondoftwo}%
\providecommand \href [0]{\begingroup \@sanitize@url \@href}%
\providecommand \@href[1]{\@@startlink{#1}\@@href}%
\providecommand \@@href[1]{\endgroup#1\@@endlink}%
\providecommand \@sanitize@url [0]{\catcode `\\12\catcode `\$12\catcode
  `\&12\catcode `\#12\catcode `\^12\catcode `\_12\catcode `\%12\relax}%
\providecommand \@@startlink[1]{}%
\providecommand \@@endlink[0]{}%
\providecommand \url  [0]{\begingroup\@sanitize@url \@url }%
\providecommand \@url [1]{\endgroup\@href {#1}{\urlprefix }}%
\providecommand \urlprefix  [0]{URL }%
\providecommand \Eprint [0]{\href }%
\providecommand \doibase [0]{https://doi.org/}%
\providecommand \selectlanguage [0]{\@gobble}%
\providecommand \bibinfo  [0]{\@secondoftwo}%
\providecommand \bibfield  [0]{\@secondoftwo}%
\providecommand \translation [1]{[#1]}%
\providecommand \BibitemOpen [0]{}%
\providecommand \bibitemStop [0]{}%
\providecommand \bibitemNoStop [0]{.\EOS\space}%
\providecommand \EOS [0]{\spacefactor3000\relax}%
\providecommand \BibitemShut  [1]{\csname bibitem#1\endcsname}%
\let\auto@bib@innerbib\@empty
\bibitem [{\citenamefont {Gibson}\ \emph {et~al.}(1970)\citenamefont {Gibson},
  \citenamefont {Kimmitt},\ and\ \citenamefont {Walker}}]{Pd_germanium}%
  \BibitemOpen
  \bibfield  {author} {\bibinfo {author} {\bibfnamefont {A.~F.}\ \bibnamefont
  {Gibson}}, \bibinfo {author} {\bibfnamefont {M.~F.}\ \bibnamefont
  {Kimmitt}},\ and\ \bibinfo {author} {\bibfnamefont {A.~C.}\ \bibnamefont
  {Walker}},\ }\bibfield  {title} {\bibinfo {title} {{Photon drag in
  germanium}},\ }\href {https://doi.org/10.1063/1.1653315} {\bibfield
  {journal} {\bibinfo  {journal} {Applied Physics Letters}\ }\textbf {\bibinfo
  {volume} {17}},\ \bibinfo {pages} {75} (\bibinfo {year} {1970})}\BibitemShut
  {NoStop}%
\bibitem [{\citenamefont {Ganichev}\ \emph {et~al.}(1983)\citenamefont
  {Ganichev}, \citenamefont {Emel'yanov},\ and\ \citenamefont
  {Yaroshetskii}}]{ganichev1983drag}%
  \BibitemOpen
  \bibfield  {author} {\bibinfo {author} {\bibfnamefont {S.}~\bibnamefont
  {Ganichev}}, \bibinfo {author} {\bibfnamefont {S.}~\bibnamefont
  {Emel'yanov}},\ and\ \bibinfo {author} {\bibfnamefont {I.}~\bibnamefont
  {Yaroshetskii}},\ }\bibfield  {title} {\bibinfo {title} {Drag of carriers by
  photons in semiconductors in the far infrared and submillimeter spectral
  ranges},\ }\href@noop {} {\bibfield  {journal} {\bibinfo  {journal}
  {Semiconductors}\ }\textbf {\bibinfo {volume} {17}},\ \bibinfo {pages} {436}
  (\bibinfo {year} {1983})}\BibitemShut {NoStop}%
\bibitem [{\citenamefont {Karch}\ \emph {et~al.}(2010)\citenamefont {Karch},
  \citenamefont {Olbrich}, \citenamefont {Schmalzbauer}, \citenamefont {Zoth},
  \citenamefont {Brinsteiner}, \citenamefont {Fehrenbacher}, \citenamefont
  {Wurstbauer}, \citenamefont {Glazov}, \citenamefont {Tarasenko},
  \citenamefont {Ivchenko}, \citenamefont {Weiss}, \citenamefont {Eroms},
  \citenamefont {Yakimova}, \citenamefont {Lara-Avila}, \citenamefont
  {Kubatkin},\ and\ \citenamefont {Ganichev}}]{Ganichev_drag_graphene}%
  \BibitemOpen
  \bibfield  {author} {\bibinfo {author} {\bibfnamefont {J.}~\bibnamefont
  {Karch}}, \bibinfo {author} {\bibfnamefont {P.}~\bibnamefont {Olbrich}},
  \bibinfo {author} {\bibfnamefont {M.}~\bibnamefont {Schmalzbauer}}, \bibinfo
  {author} {\bibfnamefont {C.}~\bibnamefont {Zoth}}, \bibinfo {author}
  {\bibfnamefont {C.}~\bibnamefont {Brinsteiner}}, \bibinfo {author}
  {\bibfnamefont {M.}~\bibnamefont {Fehrenbacher}}, \bibinfo {author}
  {\bibfnamefont {U.}~\bibnamefont {Wurstbauer}}, \bibinfo {author}
  {\bibfnamefont {M.~M.}\ \bibnamefont {Glazov}}, \bibinfo {author}
  {\bibfnamefont {S.~A.}\ \bibnamefont {Tarasenko}}, \bibinfo {author}
  {\bibfnamefont {E.~L.}\ \bibnamefont {Ivchenko}}, \bibinfo {author}
  {\bibfnamefont {D.}~\bibnamefont {Weiss}}, \bibinfo {author} {\bibfnamefont
  {J.}~\bibnamefont {Eroms}}, \bibinfo {author} {\bibfnamefont
  {R.}~\bibnamefont {Yakimova}}, \bibinfo {author} {\bibfnamefont
  {S.}~\bibnamefont {Lara-Avila}}, \bibinfo {author} {\bibfnamefont
  {S.}~\bibnamefont {Kubatkin}},\ and\ \bibinfo {author} {\bibfnamefont
  {S.~D.}\ \bibnamefont {Ganichev}},\ }\bibfield  {title} {\bibinfo {title}
  {{Dynamic Hall Effect Driven by Circularly Polarized Light in a Graphene
  Layer}},\ }\href {https://doi.org/10.1103/PhysRevLett.105.227402} {\bibfield
  {journal} {\bibinfo  {journal} {Physical Review Letters}\ }\textbf {\bibinfo
  {volume} {105}},\ \bibinfo {pages} {227402} (\bibinfo {year}
  {2010})}\BibitemShut {NoStop}%
\bibitem [{\citenamefont {Stachel}\ \emph {et~al.}(2014)\citenamefont
  {Stachel}, \citenamefont {Budkin}, \citenamefont {Hagner}, \citenamefont
  {Bel'kov}, \citenamefont {Glazov}, \citenamefont {Tarasenko}, \citenamefont
  {Clowes}, \citenamefont {Ashley}, \citenamefont {Gilbertson},\ and\
  \citenamefont {Ganichev}}]{Ganichev_drag_InSb}%
  \BibitemOpen
  \bibfield  {author} {\bibinfo {author} {\bibfnamefont {S.}~\bibnamefont
  {Stachel}}, \bibinfo {author} {\bibfnamefont {G.~V.}\ \bibnamefont {Budkin}},
  \bibinfo {author} {\bibfnamefont {U.}~\bibnamefont {Hagner}}, \bibinfo
  {author} {\bibfnamefont {V.~V.}\ \bibnamefont {Bel'kov}}, \bibinfo {author}
  {\bibfnamefont {M.~M.}\ \bibnamefont {Glazov}}, \bibinfo {author}
  {\bibfnamefont {S.~A.}\ \bibnamefont {Tarasenko}}, \bibinfo {author}
  {\bibfnamefont {S.~K.}\ \bibnamefont {Clowes}}, \bibinfo {author}
  {\bibfnamefont {T.}~\bibnamefont {Ashley}}, \bibinfo {author} {\bibfnamefont
  {A.~M.}\ \bibnamefont {Gilbertson}},\ and\ \bibinfo {author} {\bibfnamefont
  {S.~D.}\ \bibnamefont {Ganichev}},\ }\bibfield  {title} {\bibinfo {title}
  {Cyclotron-resonance-assisted photon drag effect in insb/inalsb quantum wells
  excited by terahertz radiation},\ }\href
  {https://doi.org/10.1103/PhysRevB.89.115435} {\bibfield  {journal} {\bibinfo
  {journal} {Phys. Rev. B}\ }\textbf {\bibinfo {volume} {89}},\ \bibinfo
  {pages} {115435} (\bibinfo {year} {2014})}\BibitemShut {NoStop}%
\bibitem [{\citenamefont {Shalygin}\ \emph {et~al.}(2007)\citenamefont
  {Shalygin}, \citenamefont {Diehl}, \citenamefont {Hoffmann}, \citenamefont
  {Danilov}, \citenamefont {Herrle}, \citenamefont {Tarasenko}, \citenamefont
  {Schuh}, \citenamefont {Gerl}, \citenamefont {Wegscheider}, \citenamefont
  {Prettl} \emph {et~al.}}]{Ganichev_drag_110}%
  \BibitemOpen
  \bibfield  {author} {\bibinfo {author} {\bibfnamefont {V.}~\bibnamefont
  {Shalygin}}, \bibinfo {author} {\bibfnamefont {H.}~\bibnamefont {Diehl}},
  \bibinfo {author} {\bibfnamefont {C.}~\bibnamefont {Hoffmann}}, \bibinfo
  {author} {\bibfnamefont {S.}~\bibnamefont {Danilov}}, \bibinfo {author}
  {\bibfnamefont {T.}~\bibnamefont {Herrle}}, \bibinfo {author} {\bibfnamefont
  {S.~A.}\ \bibnamefont {Tarasenko}}, \bibinfo {author} {\bibfnamefont
  {D.}~\bibnamefont {Schuh}}, \bibinfo {author} {\bibfnamefont
  {C.}~\bibnamefont {Gerl}}, \bibinfo {author} {\bibfnamefont {W.}~\bibnamefont
  {Wegscheider}}, \bibinfo {author} {\bibfnamefont {W.}~\bibnamefont {Prettl}},
  \emph {et~al.},\ }\bibfield  {title} {\bibinfo {title} {Spin photocurrents
  and the circular photon drag effect in (110)-grown quantum well structures},\
  }\href
  {https://doi.org/https://link.springer.com/article/10.1134/S0021364006220097}
  {\bibfield  {journal} {\bibinfo  {journal} {JETP letters}\ }\textbf {\bibinfo
  {volume} {84}},\ \bibinfo {pages} {570} (\bibinfo {year} {2007})}\BibitemShut
  {NoStop}%
\bibitem [{\citenamefont {Entin}\ \emph {et~al.}(2010)\citenamefont {Entin},
  \citenamefont {Magarill},\ and\ \citenamefont {Shepelyansky}}]{Entin_PD}%
  \BibitemOpen
  \bibfield  {author} {\bibinfo {author} {\bibfnamefont {M.~V.}\ \bibnamefont
  {Entin}}, \bibinfo {author} {\bibfnamefont {L.~I.}\ \bibnamefont
  {Magarill}},\ and\ \bibinfo {author} {\bibfnamefont {D.~L.}\ \bibnamefont
  {Shepelyansky}},\ }\bibfield  {title} {\bibinfo {title} {Theory of resonant
  photon drag in monolayer graphene},\ }\href
  {https://doi.org/10.1103/PhysRevB.81.165441} {\bibfield  {journal} {\bibinfo
  {journal} {Phys. Rev. B}\ }\textbf {\bibinfo {volume} {81}},\ \bibinfo
  {pages} {165441} (\bibinfo {year} {2010})}\BibitemShut {NoStop}%
\bibitem [{\citenamefont {Noginova}\ \emph {et~al.}(2011)\citenamefont
  {Noginova}, \citenamefont {Yakim}, \citenamefont {Soimo}, \citenamefont
  {Gu},\ and\ \citenamefont {Noginov}}]{Noginova_plasmon_drag}%
  \BibitemOpen
  \bibfield  {author} {\bibinfo {author} {\bibfnamefont {N.}~\bibnamefont
  {Noginova}}, \bibinfo {author} {\bibfnamefont {A.~V.}\ \bibnamefont {Yakim}},
  \bibinfo {author} {\bibfnamefont {J.}~\bibnamefont {Soimo}}, \bibinfo
  {author} {\bibfnamefont {L.}~\bibnamefont {Gu}},\ and\ \bibinfo {author}
  {\bibfnamefont {M.~A.}\ \bibnamefont {Noginov}},\ }\bibfield  {title}
  {\bibinfo {title} {{Light-to-current and current-to-light coupling in
  plasmonic systems}},\ }\href {https://doi.org/10.1103/PhysRevB.84.035447}
  {\bibfield  {journal} {\bibinfo  {journal} {Physical Review B}\ }\textbf
  {\bibinfo {volume} {84}},\ \bibinfo {pages} {1} (\bibinfo {year}
  {2011})}\BibitemShut {NoStop}%
\bibitem [{\citenamefont {Popov}\ \emph {et~al.}(2015)\citenamefont {Popov},
  \citenamefont {Fateev}, \citenamefont {Ivchenko},\ and\ \citenamefont
  {Ganichev}}]{Popov_Non_centrosymmetric}%
  \BibitemOpen
  \bibfield  {author} {\bibinfo {author} {\bibfnamefont {V.~V.}\ \bibnamefont
  {Popov}}, \bibinfo {author} {\bibfnamefont {D.~V.}\ \bibnamefont {Fateev}},
  \bibinfo {author} {\bibfnamefont {E.~L.}\ \bibnamefont {Ivchenko}},\ and\
  \bibinfo {author} {\bibfnamefont {S.~D.}\ \bibnamefont {Ganichev}},\
  }\bibfield  {title} {\bibinfo {title} {Noncentrosymmetric plasmon modes and
  giant terahertz photocurrent in a two-dimensional plasmonic crystal},\ }\href
  {https://doi.org/10.1103/PhysRevB.91.235436} {\bibfield  {journal} {\bibinfo
  {journal} {Phys. Rev. B}\ }\textbf {\bibinfo {volume} {91}},\ \bibinfo
  {pages} {235436} (\bibinfo {year} {2015})}\BibitemShut {NoStop}%
\bibitem [{\citenamefont {Olbrich}\ \emph {et~al.}(2016)\citenamefont
  {Olbrich}, \citenamefont {Kamann}, \citenamefont {K{\"{o}}nig}, \citenamefont
  {Munzert}, \citenamefont {Tutsch}, \citenamefont {Eroms}, \citenamefont
  {Weiss}, \citenamefont {Liu}, \citenamefont {Golub}, \citenamefont
  {Ivchenko}, \citenamefont {Popov}, \citenamefont {Fateev}, \citenamefont
  {Mashinsky}, \citenamefont {Fromm}, \citenamefont {Seyller},\ and\
  \citenamefont {Ganichev}}]{Olbrich_Ratchet}%
  \BibitemOpen
  \bibfield  {author} {\bibinfo {author} {\bibfnamefont {P.}~\bibnamefont
  {Olbrich}}, \bibinfo {author} {\bibfnamefont {J.}~\bibnamefont {Kamann}},
  \bibinfo {author} {\bibfnamefont {M.}~\bibnamefont {K{\"{o}}nig}}, \bibinfo
  {author} {\bibfnamefont {J.}~\bibnamefont {Munzert}}, \bibinfo {author}
  {\bibfnamefont {L.}~\bibnamefont {Tutsch}}, \bibinfo {author} {\bibfnamefont
  {J.}~\bibnamefont {Eroms}}, \bibinfo {author} {\bibfnamefont
  {D.}~\bibnamefont {Weiss}}, \bibinfo {author} {\bibfnamefont {M.-H.}\
  \bibnamefont {Liu}}, \bibinfo {author} {\bibfnamefont {L.~E.}\ \bibnamefont
  {Golub}}, \bibinfo {author} {\bibfnamefont {E.~L.}\ \bibnamefont {Ivchenko}},
  \bibinfo {author} {\bibfnamefont {V.~V.}\ \bibnamefont {Popov}}, \bibinfo
  {author} {\bibfnamefont {D.~V.}\ \bibnamefont {Fateev}}, \bibinfo {author}
  {\bibfnamefont {K.~V.}\ \bibnamefont {Mashinsky}}, \bibinfo {author}
  {\bibfnamefont {F.}~\bibnamefont {Fromm}}, \bibinfo {author} {\bibfnamefont
  {T.}~\bibnamefont {Seyller}},\ and\ \bibinfo {author} {\bibfnamefont {S.~D.}\
  \bibnamefont {Ganichev}},\ }\bibfield  {title} {\bibinfo {title} {{Terahertz
  ratchet effects in graphene with a lateral superlattice}},\ }\href
  {https://doi.org/10.1103/PhysRevB.93.075422} {\bibfield  {journal} {\bibinfo
  {journal} {Physical Review B}\ }\textbf {\bibinfo {volume} {93}},\ \bibinfo
  {pages} {075422} (\bibinfo {year} {2016})},\ \Eprint
  {https://arxiv.org/abs/1510.07946} {1510.07946} \BibitemShut {NoStop}%
\bibitem [{\citenamefont {Popov}(2013)}]{Popov_plasmon_drag}%
  \BibitemOpen
  \bibfield  {author} {\bibinfo {author} {\bibfnamefont {V.~V.}\ \bibnamefont
  {Popov}},\ }\bibfield  {title} {\bibinfo {title} {{Terahertz rectification by
  periodic two-dimensional electron plasma}},\ }\href
  {https://doi.org/10.1063/1.4811706} {\bibfield  {journal} {\bibinfo
  {journal} {Applied Physics Letters}\ }\textbf {\bibinfo {volume} {102}},\
  \bibinfo {pages} {253504} (\bibinfo {year} {2013})}\BibitemShut {NoStop}%
\bibitem [{\citenamefont {Gunyaga}\ \emph {et~al.}(2023)\citenamefont
  {Gunyaga}, \citenamefont {Durnev},\ and\ \citenamefont
  {Tarasenko}}]{Durnev_structured}%
  \BibitemOpen
  \bibfield  {author} {\bibinfo {author} {\bibfnamefont {A.~A.}\ \bibnamefont
  {Gunyaga}}, \bibinfo {author} {\bibfnamefont {M.~V.}\ \bibnamefont
  {Durnev}},\ and\ \bibinfo {author} {\bibfnamefont {S.~A.}\ \bibnamefont
  {Tarasenko}},\ }\bibfield  {title} {\bibinfo {title} {Photocurrents induced
  by structured light},\ }\href {https://doi.org/10.1103/PhysRevB.108.115402}
  {\bibfield  {journal} {\bibinfo  {journal} {Phys. Rev. B}\ }\textbf {\bibinfo
  {volume} {108}},\ \bibinfo {pages} {115402} (\bibinfo {year}
  {2023})}\BibitemShut {NoStop}%
\bibitem [{\citenamefont {Satou}\ and\ \citenamefont
  {Mikhailov}(2007)}]{Satou2007}%
  \BibitemOpen
  \bibfield  {author} {\bibinfo {author} {\bibfnamefont {A.}~\bibnamefont
  {Satou}}\ and\ \bibinfo {author} {\bibfnamefont {S.~A.}\ \bibnamefont
  {Mikhailov}},\ }\bibfield  {title} {\bibinfo {title} {{Excitation of
  two-dimensional plasmon polaritons by an incident electromagnetic wave at a
  contact}},\ }\href {https://doi.org/10.1103/PhysRevB.75.045328} {\bibfield
  {journal} {\bibinfo  {journal} {Physical Review B}\ }\textbf {\bibinfo
  {volume} {75}},\ \bibinfo {pages} {1} (\bibinfo {year} {2007})}\BibitemShut
  {NoStop}%
\bibitem [{\citenamefont {Zagorodnev}\ \emph {et~al.}(2023)\citenamefont
  {Zagorodnev}, \citenamefont {Zabolotnykh}, \citenamefont {Rodionov},\ and\
  \citenamefont {Volkov}}]{Zabolotnykh_confined}%
  \BibitemOpen
  \bibfield  {author} {\bibinfo {author} {\bibfnamefont {I.~V.}\ \bibnamefont
  {Zagorodnev}}, \bibinfo {author} {\bibfnamefont {A.~A.}\ \bibnamefont
  {Zabolotnykh}}, \bibinfo {author} {\bibfnamefont {D.~A.}\ \bibnamefont
  {Rodionov}},\ and\ \bibinfo {author} {\bibfnamefont {V.~A.}\ \bibnamefont
  {Volkov}},\ }\bibfield  {title} {\bibinfo {title} {Two-dimensional plasmons
  in laterally confined 2d electron systems},\ }\href
  {https://www.mdpi.com/2079-4991/13/6/975} {\bibfield  {journal} {\bibinfo
  {journal} {Nanomaterials}\ }\textbf {\bibinfo {volume} {13}},\ \bibinfo
  {pages} {975} (\bibinfo {year} {2023})}\BibitemShut {NoStop}%
\bibitem [{\citenamefont {Duan}\ \emph {et~al.}(2022)\citenamefont {Duan},
  \citenamefont {Alfaro‐Mozaz}, \citenamefont {Taboada‐Guti{\'{e}}rrez},
  \citenamefont {Dolado}, \citenamefont {{\'{A}}lvarez‐P{\'{e}}rez},
  \citenamefont {Titova}, \citenamefont {Bylinkin}, \citenamefont
  {Tresguerres‐Mata}, \citenamefont {Mart{\'{i}}n‐S{\'{a}}nchez},
  \citenamefont {Liu}, \citenamefont {Edgar}, \citenamefont {Bandurin},
  \citenamefont {Jarillo‐Herrero}, \citenamefont {Hillenbrand}, \citenamefont
  {Nikitin},\ and\ \citenamefont
  {Alonso‐Gonz{\'{a}}lez}}]{Titova_Near_Field}%
  \BibitemOpen
  \bibfield  {author} {\bibinfo {author} {\bibfnamefont {J.}~\bibnamefont
  {Duan}}, \bibinfo {author} {\bibfnamefont {F.~J.}\ \bibnamefont
  {Alfaro‐Mozaz}}, \bibinfo {author} {\bibfnamefont {J.}~\bibnamefont
  {Taboada‐Guti{\'{e}}rrez}}, \bibinfo {author} {\bibfnamefont
  {I.}~\bibnamefont {Dolado}}, \bibinfo {author} {\bibfnamefont
  {G.}~\bibnamefont {{\'{A}}lvarez‐P{\'{e}}rez}}, \bibinfo {author}
  {\bibfnamefont {E.}~\bibnamefont {Titova}}, \bibinfo {author} {\bibfnamefont
  {A.}~\bibnamefont {Bylinkin}}, \bibinfo {author} {\bibfnamefont {A.~I.~F.}\
  \bibnamefont {Tresguerres‐Mata}}, \bibinfo {author} {\bibfnamefont
  {J.}~\bibnamefont {Mart{\'{i}}n‐S{\'{a}}nchez}}, \bibinfo {author}
  {\bibfnamefont {S.}~\bibnamefont {Liu}}, \bibinfo {author} {\bibfnamefont
  {J.~H.}\ \bibnamefont {Edgar}}, \bibinfo {author} {\bibfnamefont {D.~A.}\
  \bibnamefont {Bandurin}}, \bibinfo {author} {\bibfnamefont {P.}~\bibnamefont
  {Jarillo‐Herrero}}, \bibinfo {author} {\bibfnamefont {R.}~\bibnamefont
  {Hillenbrand}}, \bibinfo {author} {\bibfnamefont {A.~Y.}\ \bibnamefont
  {Nikitin}},\ and\ \bibinfo {author} {\bibfnamefont {P.}~\bibnamefont
  {Alonso‐Gonz{\'{a}}lez}},\ }\bibfield  {title} {\bibinfo {title} {{Active
  and Passive Tuning of Ultranarrow Resonances in Polaritonic Nanoantennas}},\
  }\href {https://onlinelibrary.wiley.com/doi/10.1002/adma.202104954}
  {\bibfield  {journal} {\bibinfo  {journal} {Advanced Materials}\ }\textbf
  {\bibinfo {volume} {34}} (\bibinfo {year} {2022})}\BibitemShut {NoStop}%
\bibitem [{\citenamefont {Woessner}\ \emph {et~al.}(2016)\citenamefont
  {Woessner}, \citenamefont {Alonso-Gonz{\'{a}}lez}, \citenamefont {Lundeberg},
  \citenamefont {Gao}, \citenamefont {Barrios-Vargas}, \citenamefont
  {Navickaite}, \citenamefont {Ma}, \citenamefont {Janner}, \citenamefont
  {Watanabe}, \citenamefont {Cummings}, \citenamefont {Taniguchi},
  \citenamefont {Pruneri}, \citenamefont {Roche}, \citenamefont
  {Jarillo-Herrero}, \citenamefont {Hone}, \citenamefont {Hillenbrand},\ and\
  \citenamefont {Koppens}}]{Woessner_Nanoscopy}%
  \BibitemOpen
  \bibfield  {author} {\bibinfo {author} {\bibfnamefont {A.}~\bibnamefont
  {Woessner}}, \bibinfo {author} {\bibfnamefont {P.}~\bibnamefont
  {Alonso-Gonz{\'{a}}lez}}, \bibinfo {author} {\bibfnamefont {M.~B.}\
  \bibnamefont {Lundeberg}}, \bibinfo {author} {\bibfnamefont {Y.}~\bibnamefont
  {Gao}}, \bibinfo {author} {\bibfnamefont {J.~E.}\ \bibnamefont
  {Barrios-Vargas}}, \bibinfo {author} {\bibfnamefont {G.}~\bibnamefont
  {Navickaite}}, \bibinfo {author} {\bibfnamefont {Q.}~\bibnamefont {Ma}},
  \bibinfo {author} {\bibfnamefont {D.}~\bibnamefont {Janner}}, \bibinfo
  {author} {\bibfnamefont {K.}~\bibnamefont {Watanabe}}, \bibinfo {author}
  {\bibfnamefont {A.~W.}\ \bibnamefont {Cummings}}, \bibinfo {author}
  {\bibfnamefont {T.}~\bibnamefont {Taniguchi}}, \bibinfo {author}
  {\bibfnamefont {V.}~\bibnamefont {Pruneri}}, \bibinfo {author} {\bibfnamefont
  {S.}~\bibnamefont {Roche}}, \bibinfo {author} {\bibfnamefont
  {P.}~\bibnamefont {Jarillo-Herrero}}, \bibinfo {author} {\bibfnamefont
  {J.}~\bibnamefont {Hone}}, \bibinfo {author} {\bibfnamefont {R.}~\bibnamefont
  {Hillenbrand}},\ and\ \bibinfo {author} {\bibfnamefont {F.~H.~L.}\
  \bibnamefont {Koppens}},\ }\bibfield  {title} {\bibinfo {title} {{Near-field
  photocurrent nanoscopy on bare and encapsulated graphene}},\ }\href
  {https://doi.org/10.1038/ncomms10783} {\bibfield  {journal} {\bibinfo
  {journal} {Nature Communications}\ }\textbf {\bibinfo {volume} {7}},\
  \bibinfo {pages} {10783} (\bibinfo {year} {2016})},\ \Eprint
  {https://arxiv.org/abs/1508.07864} {1508.07864} \BibitemShut {NoStop}%
\bibitem [{\citenamefont {Soltani}\ \emph {et~al.}(2020)\citenamefont
  {Soltani}, \citenamefont {Kuschewski}, \citenamefont {Bonmann}, \citenamefont
  {Generalov}, \citenamefont {Vorobiev}, \citenamefont {Ludwig}, \citenamefont
  {Wiecha}, \citenamefont {Cibiraite}, \citenamefont {Walla}, \citenamefont
  {Winnerl}, \citenamefont {Kehr}, \citenamefont {Eng}, \citenamefont {Stake},\
  and\ \citenamefont {Roskos}}]{Roskos_SNOM}%
  \BibitemOpen
  \bibfield  {author} {\bibinfo {author} {\bibfnamefont {A.}~\bibnamefont
  {Soltani}}, \bibinfo {author} {\bibfnamefont {F.}~\bibnamefont {Kuschewski}},
  \bibinfo {author} {\bibfnamefont {M.}~\bibnamefont {Bonmann}}, \bibinfo
  {author} {\bibfnamefont {A.}~\bibnamefont {Generalov}}, \bibinfo {author}
  {\bibfnamefont {A.}~\bibnamefont {Vorobiev}}, \bibinfo {author}
  {\bibfnamefont {F.}~\bibnamefont {Ludwig}}, \bibinfo {author} {\bibfnamefont
  {M.~M.}\ \bibnamefont {Wiecha}}, \bibinfo {author} {\bibfnamefont
  {D.}~\bibnamefont {Cibiraite}}, \bibinfo {author} {\bibfnamefont
  {F.}~\bibnamefont {Walla}}, \bibinfo {author} {\bibfnamefont
  {S.}~\bibnamefont {Winnerl}}, \bibinfo {author} {\bibfnamefont {S.~C.}\
  \bibnamefont {Kehr}}, \bibinfo {author} {\bibfnamefont {L.~M.}\ \bibnamefont
  {Eng}}, \bibinfo {author} {\bibfnamefont {J.}~\bibnamefont {Stake}},\ and\
  \bibinfo {author} {\bibfnamefont {H.~G.}\ \bibnamefont {Roskos}},\ }\bibfield
   {title} {\bibinfo {title} {{Direct nanoscopic observation of plasma waves in
  the channel of a graphene field-effect transistor}},\ }\href
  {https://doi.org/10.1038/s41377-020-0321-0} {\bibfield  {journal} {\bibinfo
  {journal} {Light: Science Applications}\ }\textbf {\bibinfo {volume} {9}},\
  \bibinfo {pages} {97} (\bibinfo {year} {2020})}\BibitemShut {NoStop}%
\bibitem [{\citenamefont {Alonso-Gonzalez}\ \emph {et~al.}(2014)\citenamefont
  {Alonso-Gonzalez}, \citenamefont {Nikitin}, \citenamefont {Golmar},
  \citenamefont {Centeno}, \citenamefont {Pesquera}, \citenamefont {Velez},
  \citenamefont {Chen}, \citenamefont {Navickaite}, \citenamefont {Koppens},
  \citenamefont {Zurutuza}, \citenamefont {Casanova}, \citenamefont {Hueso},\
  and\ \citenamefont {Hillenbrand}}]{Alonso-Gonzalez2014}%
  \BibitemOpen
  \bibfield  {author} {\bibinfo {author} {\bibfnamefont {P.}~\bibnamefont
  {Alonso-Gonzalez}}, \bibinfo {author} {\bibfnamefont {A.~Y.}\ \bibnamefont
  {Nikitin}}, \bibinfo {author} {\bibfnamefont {F.}~\bibnamefont {Golmar}},
  \bibinfo {author} {\bibfnamefont {A.}~\bibnamefont {Centeno}}, \bibinfo
  {author} {\bibfnamefont {A.}~\bibnamefont {Pesquera}}, \bibinfo {author}
  {\bibfnamefont {S.}~\bibnamefont {Velez}}, \bibinfo {author} {\bibfnamefont
  {J.}~\bibnamefont {Chen}}, \bibinfo {author} {\bibfnamefont {G.}~\bibnamefont
  {Navickaite}}, \bibinfo {author} {\bibfnamefont {F.}~\bibnamefont {Koppens}},
  \bibinfo {author} {\bibfnamefont {A.}~\bibnamefont {Zurutuza}}, \bibinfo
  {author} {\bibfnamefont {F.}~\bibnamefont {Casanova}}, \bibinfo {author}
  {\bibfnamefont {L.~E.}\ \bibnamefont {Hueso}},\ and\ \bibinfo {author}
  {\bibfnamefont {R.}~\bibnamefont {Hillenbrand}},\ }\bibfield  {title}
  {\bibinfo {title} {{Controlling graphene plasmons with resonant metal
  antennas and spatial conductivity patterns}},\ }\href
  {https://doi.org/10.1126/science.1253202} {\bibfield  {journal} {\bibinfo
  {journal} {Science}\ }\textbf {\bibinfo {volume} {344}},\ \bibinfo {pages}
  {1369} (\bibinfo {year} {2014})}\BibitemShut {NoStop}%
\bibitem [{\citenamefont {Ni}\ \emph {et~al.}(2018)\citenamefont {Ni},
  \citenamefont {McLeod}, \citenamefont {Sun}, \citenamefont {Wang},
  \citenamefont {Xiong}, \citenamefont {Post}, \citenamefont {Sunku},
  \citenamefont {Jiang}, \citenamefont {Hone}, \citenamefont {Dean},
  \citenamefont {Fogler},\ and\ \citenamefont {Basov}}]{Ni2018}%
  \BibitemOpen
  \bibfield  {author} {\bibinfo {author} {\bibfnamefont {G.~X.}\ \bibnamefont
  {Ni}}, \bibinfo {author} {\bibfnamefont {A.~S.}\ \bibnamefont {McLeod}},
  \bibinfo {author} {\bibfnamefont {Z.}~\bibnamefont {Sun}}, \bibinfo {author}
  {\bibfnamefont {L.}~\bibnamefont {Wang}}, \bibinfo {author} {\bibfnamefont
  {L.}~\bibnamefont {Xiong}}, \bibinfo {author} {\bibfnamefont {K.~W.}\
  \bibnamefont {Post}}, \bibinfo {author} {\bibfnamefont {S.~S.}\ \bibnamefont
  {Sunku}}, \bibinfo {author} {\bibfnamefont {B.-Y.}\ \bibnamefont {Jiang}},
  \bibinfo {author} {\bibfnamefont {J.}~\bibnamefont {Hone}}, \bibinfo {author}
  {\bibfnamefont {C.~R.}\ \bibnamefont {Dean}}, \bibinfo {author}
  {\bibfnamefont {M.~M.}\ \bibnamefont {Fogler}},\ and\ \bibinfo {author}
  {\bibfnamefont {D.~N.}\ \bibnamefont {Basov}},\ }\bibfield  {title} {\bibinfo
  {title} {{Fundamental limits to graphene plasmonics}},\ }\href
  {https://doi.org/10.1038/s41586-018-0136-9} {\bibfield  {journal} {\bibinfo
  {journal} {Nature}\ }\textbf {\bibinfo {volume} {557}},\ \bibinfo {pages}
  {530} (\bibinfo {year} {2018})}\BibitemShut {NoStop}%
\bibitem [{\citenamefont {Tielrooij}\ \emph
  {et~al.}(2015{\natexlab{a}})\citenamefont {Tielrooij}, \citenamefont
  {Massicotte}, \citenamefont {Piatkowski}, \citenamefont {Woessner},
  \citenamefont {Ma}, \citenamefont {Jarillo-Herrero}, \citenamefont {van
  Hulst},\ and\ \citenamefont {Koppens}}]{Tielrooij_Photocurrent}%
  \BibitemOpen
  \bibfield  {author} {\bibinfo {author} {\bibfnamefont {K.~J.}\ \bibnamefont
  {Tielrooij}}, \bibinfo {author} {\bibfnamefont {M.}~\bibnamefont
  {Massicotte}}, \bibinfo {author} {\bibfnamefont {L.}~\bibnamefont
  {Piatkowski}}, \bibinfo {author} {\bibfnamefont {A.}~\bibnamefont
  {Woessner}}, \bibinfo {author} {\bibfnamefont {Q.}~\bibnamefont {Ma}},
  \bibinfo {author} {\bibfnamefont {P.}~\bibnamefont {Jarillo-Herrero}},
  \bibinfo {author} {\bibfnamefont {N.~F.}\ \bibnamefont {van Hulst}},\ and\
  \bibinfo {author} {\bibfnamefont {F.~H.~L.}\ \bibnamefont {Koppens}},\
  }\bibfield  {title} {\bibinfo {title} {{Hot-carrier photocurrent effects at
  graphene–metal interfaces}},\ }\href
  {https://doi.org/10.1088/0953-8984/27/16/164207} {\bibfield  {journal}
  {\bibinfo  {journal} {Journal of Physics: Condensed Matter}\ }\textbf
  {\bibinfo {volume} {27}},\ \bibinfo {pages} {164207} (\bibinfo {year}
  {2015}{\natexlab{a}})},\ \Eprint {https://arxiv.org/abs/1411.5665}
  {1411.5665} \BibitemShut {NoStop}%
\bibitem [{\citenamefont {Tielrooij}\ \emph
  {et~al.}(2015{\natexlab{b}})\citenamefont {Tielrooij}, \citenamefont
  {Piatkowski}, \citenamefont {Massicotte}, \citenamefont {Woessner},
  \citenamefont {Ma}, \citenamefont {Lee}, \citenamefont {Myhro}, \citenamefont
  {Lau}, \citenamefont {Jarillo-Herrero}, \citenamefont {van Hulst},\ and\
  \citenamefont {Koppens}}]{Tielrooij_Photocurrent_NNano}%
  \BibitemOpen
  \bibfield  {author} {\bibinfo {author} {\bibfnamefont {K.~J.}\ \bibnamefont
  {Tielrooij}}, \bibinfo {author} {\bibfnamefont {L.}~\bibnamefont
  {Piatkowski}}, \bibinfo {author} {\bibfnamefont {M.}~\bibnamefont
  {Massicotte}}, \bibinfo {author} {\bibfnamefont {A.}~\bibnamefont
  {Woessner}}, \bibinfo {author} {\bibfnamefont {Q.}~\bibnamefont {Ma}},
  \bibinfo {author} {\bibfnamefont {Y.}~\bibnamefont {Lee}}, \bibinfo {author}
  {\bibfnamefont {K.~S.}\ \bibnamefont {Myhro}}, \bibinfo {author}
  {\bibfnamefont {C.~N.}\ \bibnamefont {Lau}}, \bibinfo {author} {\bibfnamefont
  {P.}~\bibnamefont {Jarillo-Herrero}}, \bibinfo {author} {\bibfnamefont
  {N.~F.}\ \bibnamefont {van Hulst}},\ and\ \bibinfo {author} {\bibfnamefont
  {F.~H.~L.}\ \bibnamefont {Koppens}},\ }\bibfield  {title} {\bibinfo {title}
  {{Generation of photovoltage in graphene on a femtosecond timescale through
  efficient carrier heating}},\ }\href {https://doi.org/10.1038/nnano.2015.54}
  {\bibfield  {journal} {\bibinfo  {journal} {Nature Nanotechnology}\ }\textbf
  {\bibinfo {volume} {10}},\ \bibinfo {pages} {437} (\bibinfo {year}
  {2015}{\natexlab{b}})},\ \Eprint {https://arxiv.org/abs/1504.06487}
  {1504.06487} \BibitemShut {NoStop}%
\bibitem [{\citenamefont {Candussio}\ \emph {et~al.}(2020)\citenamefont
  {Candussio}, \citenamefont {Durnev}, \citenamefont {Tarasenko}, \citenamefont
  {Yin}, \citenamefont {Keil}, \citenamefont {Yang}, \citenamefont {Son},
  \citenamefont {Mishchenko}, \citenamefont {Plank}, \citenamefont {Bel'kov},
  \citenamefont {Slizovskiy}, \citenamefont {Fal'ko},\ and\ \citenamefont
  {Ganichev}}]{Ganichev_BLG_subwavelength}%
  \BibitemOpen
  \bibfield  {author} {\bibinfo {author} {\bibfnamefont {S.}~\bibnamefont
  {Candussio}}, \bibinfo {author} {\bibfnamefont {M.~V.}\ \bibnamefont
  {Durnev}}, \bibinfo {author} {\bibfnamefont {S.~A.}\ \bibnamefont
  {Tarasenko}}, \bibinfo {author} {\bibfnamefont {J.}~\bibnamefont {Yin}},
  \bibinfo {author} {\bibfnamefont {J.}~\bibnamefont {Keil}}, \bibinfo {author}
  {\bibfnamefont {Y.}~\bibnamefont {Yang}}, \bibinfo {author} {\bibfnamefont
  {S.-K.}\ \bibnamefont {Son}}, \bibinfo {author} {\bibfnamefont
  {A.}~\bibnamefont {Mishchenko}}, \bibinfo {author} {\bibfnamefont
  {H.}~\bibnamefont {Plank}}, \bibinfo {author} {\bibfnamefont {V.~V.}\
  \bibnamefont {Bel'kov}}, \bibinfo {author} {\bibfnamefont {S.}~\bibnamefont
  {Slizovskiy}}, \bibinfo {author} {\bibfnamefont {V.}~\bibnamefont {Fal'ko}},\
  and\ \bibinfo {author} {\bibfnamefont {S.~D.}\ \bibnamefont {Ganichev}},\
  }\bibfield  {title} {\bibinfo {title} {Edge photocurrent driven by terahertz
  electric field in bilayer graphene},\ }\href
  {https://doi.org/10.1103/PhysRevB.102.045406} {\bibfield  {journal} {\bibinfo
   {journal} {Phys. Rev. B}\ }\textbf {\bibinfo {volume} {102}},\ \bibinfo
  {pages} {045406} (\bibinfo {year} {2020})}\BibitemShut {NoStop}%
\bibitem [{\citenamefont {Sommerfeld}(1896)}]{sommerfeld1896}%
  \BibitemOpen
  \bibfield  {author} {\bibinfo {author} {\bibfnamefont {A.}~\bibnamefont
  {Sommerfeld}},\ }\bibfield  {title} {\bibinfo {title} {Mathematische theorie
  der diffraction},\ }\href
  {https://link.springer.com/article/10.1007/BF01447273} {\bibfield  {journal}
  {\bibinfo  {journal} {Mathematische Annalen}\ }\textbf {\bibinfo {volume}
  {47}},\ \bibinfo {pages} {317} (\bibinfo {year} {1896})}\BibitemShut
  {NoStop}%
\bibitem [{\citenamefont {Landau}\ and\ \citenamefont
  {Lifshitz}(2013)}]{landau2013electrodynamics}%
  \BibitemOpen
  \bibfield  {author} {\bibinfo {author} {\bibfnamefont {L.~D.}\ \bibnamefont
  {Landau}}\ and\ \bibinfo {author} {\bibfnamefont {E.~M.}\ \bibnamefont
  {Lifshitz}},\ }\href@noop {} {\emph {\bibinfo {title} {Electrodynamics of
  continuous media}}},\ Vol.~\bibinfo {volume} {8}\ (\bibinfo  {publisher}
  {Elsevier},\ \bibinfo {year} {2013})\ \bibinfo {note} {see chapter 3
  ''Methods for electroststic problem solutions'' and chapter 94 ''Diffraction
  at the wedge'' for the limiting behaviour of electric field upon scattering
  at the acute objects}\BibitemShut {NoStop}%
\bibitem [{\citenamefont {Vainstein}(1966)}]{Vainstein}%
  \BibitemOpen
  \bibfield  {author} {\bibinfo {author} {\bibfnamefont {L.}~\bibnamefont
  {Vainstein}},\ }\href@noop {} {\emph {\bibinfo {title} {Theory of diffraction
  and method of factorization}}}\ (\bibinfo  {publisher} {Soviet Radio},\
  \bibinfo {year} {1966})\ \bibinfo {note} {in Russian}\BibitemShut {NoStop}%
\bibitem [{\citenamefont {Daniele}\ and\ \citenamefont
  {Zich}(2014)}]{daniele2014wiener}%
  \BibitemOpen
  \bibfield  {author} {\bibinfo {author} {\bibfnamefont {V.~G.}\ \bibnamefont
  {Daniele}}\ and\ \bibinfo {author} {\bibfnamefont {R.}~\bibnamefont {Zich}},\
  }\href@noop {} {\emph {\bibinfo {title} {The {W}iener-{H}opf method in
  electromagnetics}}}\ (\bibinfo  {publisher} {SciTech Publishing
  Incorporated},\ \bibinfo {year} {2014})\BibitemShut {NoStop}%
\bibitem [{\citenamefont {Dyakonov}\ and\ \citenamefont
  {Shur}(1996)}]{DS_rectification}%
  \BibitemOpen
  \bibfield  {author} {\bibinfo {author} {\bibfnamefont {M.}~\bibnamefont
  {Dyakonov}}\ and\ \bibinfo {author} {\bibfnamefont {M.}~\bibnamefont
  {Shur}},\ }\bibfield  {title} {\bibinfo {title} {{Detection, mixing, and
  frequency multiplication of terahertz radiation by two-dimensional electronic
  fluid}},\ }\href {https://doi.org/10.1109/16.485650} {\bibfield  {journal}
  {\bibinfo  {journal} {IEEE Transactions on Electron Devices}\ }\textbf
  {\bibinfo {volume} {43}},\ \bibinfo {pages} {380} (\bibinfo {year}
  {1996})}\BibitemShut {NoStop}%
\bibitem [{\citenamefont {Sakowicz}\ \emph {et~al.}(2011)\citenamefont
  {Sakowicz}, \citenamefont {Lifshits}, \citenamefont {Klimenko}, \citenamefont
  {Schuster}, \citenamefont {Coquillat}, \citenamefont {Teppe},\ and\
  \citenamefont {Knap}}]{Sakowicz2011}%
  \BibitemOpen
  \bibfield  {author} {\bibinfo {author} {\bibfnamefont {M.}~\bibnamefont
  {Sakowicz}}, \bibinfo {author} {\bibfnamefont {M.~B.}\ \bibnamefont
  {Lifshits}}, \bibinfo {author} {\bibfnamefont {O.~A.}\ \bibnamefont
  {Klimenko}}, \bibinfo {author} {\bibfnamefont {F.}~\bibnamefont {Schuster}},
  \bibinfo {author} {\bibfnamefont {D.}~\bibnamefont {Coquillat}}, \bibinfo
  {author} {\bibfnamefont {F.}~\bibnamefont {Teppe}},\ and\ \bibinfo {author}
  {\bibfnamefont {W.}~\bibnamefont {Knap}},\ }\bibfield  {title} {\bibinfo
  {title} {{Terahertz responsivity of field effect transistors versus their
  static channel conductivity and loading effects}},\ }\href
  {https://doi.org/10.1063/1.3632058} {\bibfield  {journal} {\bibinfo
  {journal} {Journal of Applied Physics}\ }\textbf {\bibinfo {volume} {110}},\
  \bibinfo {pages} {054512} (\bibinfo {year} {2011})}\BibitemShut {NoStop}%
\bibitem [{\citenamefont {Lisauskas}\ \emph {et~al.}(2009)\citenamefont
  {Lisauskas}, \citenamefont {Pfeiffer}, \citenamefont {Öjefors},
  \citenamefont {Bolìvar}, \citenamefont {Glaab},\ and\ \citenamefont
  {Roskos}}]{Roskos_DRSM}%
  \BibitemOpen
  \bibfield  {author} {\bibinfo {author} {\bibfnamefont {A.}~\bibnamefont
  {Lisauskas}}, \bibinfo {author} {\bibfnamefont {U.}~\bibnamefont {Pfeiffer}},
  \bibinfo {author} {\bibfnamefont {E.}~\bibnamefont {Öjefors}}, \bibinfo
  {author} {\bibfnamefont {P.~H.}\ \bibnamefont {Bolìvar}}, \bibinfo {author}
  {\bibfnamefont {D.}~\bibnamefont {Glaab}},\ and\ \bibinfo {author}
  {\bibfnamefont {H.~G.}\ \bibnamefont {Roskos}},\ }\bibfield  {title}
  {\bibinfo {title} {{Rational design of high-responsivity detectors of
  terahertz radiation based on distributed self-mixing in silicon field-effect
  transistors}},\ }\href {https://doi.org/10.1063/1.3140611} {\bibfield
  {journal} {\bibinfo  {journal} {Journal of Applied Physics}\ }\textbf
  {\bibinfo {volume} {105}},\ \bibinfo {pages} {114511} (\bibinfo {year}
  {2009})}\BibitemShut {NoStop}%
\bibitem [{\citenamefont {Matyushkin}\ \emph {et~al.}(2020)\citenamefont
  {Matyushkin}, \citenamefont {Danilov}, \citenamefont {Moskotin},
  \citenamefont {Belosevich}, \citenamefont {Kaurova}, \citenamefont {Rybin},
  \citenamefont {Obraztsova}, \citenamefont {Fedorov}, \citenamefont
  {Gorbenko}, \citenamefont {Kachorovskii},\ and\ \citenamefont
  {Ganichev}}]{Matyushkin2020}%
  \BibitemOpen
  \bibfield  {author} {\bibinfo {author} {\bibfnamefont {Y.}~\bibnamefont
  {Matyushkin}}, \bibinfo {author} {\bibfnamefont {S.}~\bibnamefont {Danilov}},
  \bibinfo {author} {\bibfnamefont {M.}~\bibnamefont {Moskotin}}, \bibinfo
  {author} {\bibfnamefont {V.}~\bibnamefont {Belosevich}}, \bibinfo {author}
  {\bibfnamefont {N.}~\bibnamefont {Kaurova}}, \bibinfo {author} {\bibfnamefont
  {M.}~\bibnamefont {Rybin}}, \bibinfo {author} {\bibfnamefont {E.~D.}\
  \bibnamefont {Obraztsova}}, \bibinfo {author} {\bibfnamefont
  {G.}~\bibnamefont {Fedorov}}, \bibinfo {author} {\bibfnamefont
  {I.}~\bibnamefont {Gorbenko}}, \bibinfo {author} {\bibfnamefont
  {V.}~\bibnamefont {Kachorovskii}},\ and\ \bibinfo {author} {\bibfnamefont
  {S.}~\bibnamefont {Ganichev}},\ }\bibfield  {title} {\bibinfo {title}
  {{Helicity-Sensitive Plasmonic Terahertz Interferometer}},\ }\href
  {https://doi.org/10.1021/acs.nanolett.0c02692} {\bibfield  {journal}
  {\bibinfo  {journal} {Nano Letters}\ }\textbf {\bibinfo {volume} {20}},\
  \bibinfo {pages} {7296} (\bibinfo {year} {2020})},\ \Eprint
  {https://arxiv.org/abs/2007.01035} {2007.01035} \BibitemShut {NoStop}%
\bibitem [{\citenamefont {Tomadin}\ and\ \citenamefont
  {Polini}(2013)}]{Polini_PW_rectification}%
  \BibitemOpen
  \bibfield  {author} {\bibinfo {author} {\bibfnamefont {A.}~\bibnamefont
  {Tomadin}}\ and\ \bibinfo {author} {\bibfnamefont {M.}~\bibnamefont
  {Polini}},\ }\bibfield  {title} {\bibinfo {title} {Theory of the plasma-wave
  photoresponse of a gated graphene sheet},\ }\href
  {https://doi.org/10.1103/PhysRevB.88.205426} {\bibfield  {journal} {\bibinfo
  {journal} {Phys. Rev. B}\ }\textbf {\bibinfo {volume} {88}},\ \bibinfo
  {pages} {205426} (\bibinfo {year} {2013})}\BibitemShut {NoStop}%
\bibitem [{\citenamefont {Fateev}\ \emph {et~al.}(2017)\citenamefont {Fateev},
  \citenamefont {Mashinsky},\ and\ \citenamefont
  {Popov}}]{Fateev_Rectification}%
  \BibitemOpen
  \bibfield  {author} {\bibinfo {author} {\bibfnamefont {D.~V.}\ \bibnamefont
  {Fateev}}, \bibinfo {author} {\bibfnamefont {K.~V.}\ \bibnamefont
  {Mashinsky}},\ and\ \bibinfo {author} {\bibfnamefont {V.~V.}\ \bibnamefont
  {Popov}},\ }\bibfield  {title} {\bibinfo {title} {{Terahertz plasmonic
  rectification in a spatially periodic graphene}},\ }\href
  {https://doi.org/10.1063/1.4975829} {\bibfield  {journal} {\bibinfo
  {journal} {Applied Physics Letters}\ }\textbf {\bibinfo {volume} {110}},\
  \bibinfo {pages} {061106} (\bibinfo {year} {2017})}\BibitemShut {NoStop}%
\bibitem [{\citenamefont {Ludwig}\ \emph {et~al.}()\citenamefont {Ludwig},
  \citenamefont {Roskos},\ and\ \citenamefont {Borsche}}]{ludwig20242d}%
  \BibitemOpen
  \bibfield  {author} {\bibinfo {author} {\bibfnamefont {F.}~\bibnamefont
  {Ludwig}}, \bibinfo {author} {\bibfnamefont {H.~G.}\ \bibnamefont {Roskos}},\
  and\ \bibinfo {author} {\bibfnamefont {R.}~\bibnamefont {Borsche}},\
  }\bibfield  {title} {\bibinfo {title} {2d hydrodynamic simulation of
  {T}era{F}{E}{T}s beyond the gradual-channel approximation for transient,
  large-signal or ultrahigh-frequency simulations},\ }\href
  {https://arxiv.org/abs/2405.18764} {\bibinfo  {journal} {arXiv preprint
  arXiv:2405.18764}\ }\BibitemShut {NoStop}%
\bibitem [{\citenamefont {Nikulin}\ \emph {et~al.}(2021)\citenamefont
  {Nikulin}, \citenamefont {Mylnikov}, \citenamefont {Bandurin},\ and\
  \citenamefont {Svintsov}}]{Nikulin_Edge}%
  \BibitemOpen
\bibfield  {journal} {  }\bibfield  {author} {\bibinfo {author} {\bibfnamefont
  {E.}~\bibnamefont {Nikulin}}, \bibinfo {author} {\bibfnamefont
  {D.}~\bibnamefont {Mylnikov}}, \bibinfo {author} {\bibfnamefont
  {D.}~\bibnamefont {Bandurin}},\ and\ \bibinfo {author} {\bibfnamefont
  {D.}~\bibnamefont {Svintsov}},\ }\bibfield  {title} {\bibinfo {title} {{Edge
  diffraction, plasmon launching, and universal absorption enhancement in
  two-dimensional junctions}},\ }\href
  {https://doi.org/10.1103/PhysRevB.103.085306} {\bibfield  {journal} {\bibinfo
   {journal} {Physical Review B}\ }\textbf {\bibinfo {volume} {103}},\ \bibinfo
  {pages} {085306} (\bibinfo {year} {2021})}\BibitemShut {NoStop}%
\bibitem [{\citenamefont {Svintsov}(2024)}]{svintsov2024exact}%
  \BibitemOpen
  \bibfield  {author} {\bibinfo {author} {\bibfnamefont {D.}~\bibnamefont
  {Svintsov}},\ }\bibfield  {title} {\bibinfo {title} {Exact theory of edge
  diffraction and launching of transverse electric plasmons at two-dimensional
  junctions},\ }\href {https://arxiv.org/abs/2410.20206} {\bibfield  {journal}
  {\bibinfo  {journal} {arXiv preprint arXiv:2410.20206}\ } (\bibinfo {year}
  {2024})}\BibitemShut {NoStop}%
\bibitem [{\citenamefont {M{\"{o}}nch}\ \emph {et~al.}(2022)\citenamefont
  {M{\"{o}}nch}, \citenamefont {Potashin}, \citenamefont {Lindner},
  \citenamefont {Yahniuk}, \citenamefont {Golub}, \citenamefont {Kachorovskii},
  \citenamefont {Bel'kov}, \citenamefont {Huber}, \citenamefont {Watanabe},
  \citenamefont {Taniguchi}, \citenamefont {Eroms}, \citenamefont {Weiss},\
  and\ \citenamefont {Ganichev}}]{Ratchet_signature}%
  \BibitemOpen
  \bibfield  {author} {\bibinfo {author} {\bibfnamefont {E.}~\bibnamefont
  {M{\"{o}}nch}}, \bibinfo {author} {\bibfnamefont {S.~O.}\ \bibnamefont
  {Potashin}}, \bibinfo {author} {\bibfnamefont {K.}~\bibnamefont {Lindner}},
  \bibinfo {author} {\bibfnamefont {I.}~\bibnamefont {Yahniuk}}, \bibinfo
  {author} {\bibfnamefont {L.~E.}\ \bibnamefont {Golub}}, \bibinfo {author}
  {\bibfnamefont {V.~Y.}\ \bibnamefont {Kachorovskii}}, \bibinfo {author}
  {\bibfnamefont {V.~V.}\ \bibnamefont {Bel'kov}}, \bibinfo {author}
  {\bibfnamefont {R.}~\bibnamefont {Huber}}, \bibinfo {author} {\bibfnamefont
  {K.}~\bibnamefont {Watanabe}}, \bibinfo {author} {\bibfnamefont
  {T.}~\bibnamefont {Taniguchi}}, \bibinfo {author} {\bibfnamefont
  {J.}~\bibnamefont {Eroms}}, \bibinfo {author} {\bibfnamefont
  {D.}~\bibnamefont {Weiss}},\ and\ \bibinfo {author} {\bibfnamefont {S.~D.}\
  \bibnamefont {Ganichev}},\ }\bibfield  {title} {\bibinfo {title} {{Ratchet
  effect in spatially modulated bilayer graphene: Signature of hydrodynamic
  transport}},\ }\href {https://doi.org/10.1103/PhysRevB.105.045404} {\bibfield
   {journal} {\bibinfo  {journal} {Physical Review B}\ }\textbf {\bibinfo
  {volume} {105}},\ \bibinfo {pages} {045404} (\bibinfo {year}
  {2022})}\BibitemShut {NoStop}%
\bibitem [{\citenamefont {Ma}\ \emph {et~al.}(2014)\citenamefont {Ma},
  \citenamefont {Gabor}, \citenamefont {Andersen}, \citenamefont {Nair},
  \citenamefont {Watanabe}, \citenamefont {Taniguchi},\ and\ \citenamefont
  {Jarillo-Herrero}}]{Herrero_TE}%
  \BibitemOpen
  \bibfield  {author} {\bibinfo {author} {\bibfnamefont {Q.}~\bibnamefont
  {Ma}}, \bibinfo {author} {\bibfnamefont {N.~M.}\ \bibnamefont {Gabor}},
  \bibinfo {author} {\bibfnamefont {T.~I.}\ \bibnamefont {Andersen}}, \bibinfo
  {author} {\bibfnamefont {N.~L.}\ \bibnamefont {Nair}}, \bibinfo {author}
  {\bibfnamefont {K.}~\bibnamefont {Watanabe}}, \bibinfo {author}
  {\bibfnamefont {T.}~\bibnamefont {Taniguchi}},\ and\ \bibinfo {author}
  {\bibfnamefont {P.}~\bibnamefont {Jarillo-Herrero}},\ }\bibfield  {title}
  {\bibinfo {title} {{Competing Channels for Hot-Electron Cooling in
  Graphene}},\ }\href {https://doi.org/10.1103/PhysRevLett.112.247401}
  {\bibfield  {journal} {\bibinfo  {journal} {Physical Review Letters}\
  }\textbf {\bibinfo {volume} {112}},\ \bibinfo {pages} {247401} (\bibinfo
  {year} {2014})}\BibitemShut {NoStop}%
\bibitem [{\citenamefont {Gabor}\ \emph {et~al.}(2011)\citenamefont {Gabor},
  \citenamefont {Song}, \citenamefont {Ma}, \citenamefont {Nair}, \citenamefont
  {Taychatanapat}, \citenamefont {Watanabe}, \citenamefont {Taniguchi},
  \citenamefont {Levitov},\ and\ \citenamefont
  {Jarillo-Herrero}}]{Levitov_PTE}%
  \BibitemOpen
  \bibfield  {author} {\bibinfo {author} {\bibfnamefont {N.~M.}\ \bibnamefont
  {Gabor}}, \bibinfo {author} {\bibfnamefont {J.~C.~W.}\ \bibnamefont {Song}},
  \bibinfo {author} {\bibfnamefont {Q.}~\bibnamefont {Ma}}, \bibinfo {author}
  {\bibfnamefont {N.~L.}\ \bibnamefont {Nair}}, \bibinfo {author}
  {\bibfnamefont {T.}~\bibnamefont {Taychatanapat}}, \bibinfo {author}
  {\bibfnamefont {K.}~\bibnamefont {Watanabe}}, \bibinfo {author}
  {\bibfnamefont {T.}~\bibnamefont {Taniguchi}}, \bibinfo {author}
  {\bibfnamefont {L.~S.}\ \bibnamefont {Levitov}},\ and\ \bibinfo {author}
  {\bibfnamefont {P.}~\bibnamefont {Jarillo-Herrero}},\ }\bibfield  {title}
  {\bibinfo {title} {{Hot Carrier–Assisted Intrinsic Photoresponse in
  Graphene}},\ }\href {https://doi.org/10.1126/science.1211384} {\bibfield
  {journal} {\bibinfo  {journal} {Science}\ }\textbf {\bibinfo {volume}
  {334}},\ \bibinfo {pages} {648} (\bibinfo {year} {2011})}\BibitemShut
  {NoStop}%
\bibitem [{\citenamefont {Ivchenko}(2014)}]{Ivchenko_heating_in_FETs}%
  \BibitemOpen
  \bibfield  {author} {\bibinfo {author} {\bibfnamefont {E.~L.}\ \bibnamefont
  {Ivchenko}},\ }\bibfield  {title} {\bibinfo {title} {{Effect of carrier
  heating on photovoltage in FET}},\ }\href
  {https://doi.org/10.1134/S1063783414120142} {\bibfield  {journal} {\bibinfo
  {journal} {Physics of the Solid State}\ }\textbf {\bibinfo {volume} {56}},\
  \bibinfo {pages} {2514} (\bibinfo {year} {2014})}\BibitemShut {NoStop}%
\bibitem [{Note1()}]{Note1}%
  \BibitemOpen
  \bibinfo {note} {To prove this, we denote $I = \DOTSI \intop \ilimits@
  _{-\infty }^{+\infty }{E^* \partial _x E dx}$. The expression for photon drag
  voltage across the extended 2DES is proportional to ${\protect \rm Im}
  (\sigma _\omega I) = {\protect \rm Im} \sigma _\omega {\protect \rm Re} I +
  {\protect \rm Re} \sigma _\omega {\protect \rm Im} I$. To show that the first
  term is zero, we write ${\protect \rm Re} I = (\DOTSI \intop \ilimits@
  _{-\infty }^{+\infty }{E^* \partial _x E dx} + \DOTSI \intop \ilimits@
  _{-\infty }^{+\infty }{E \partial _x E^* dx})/2$. Integrating the second
  summand by parts, we get ${\protect \rm Re} I = |E|^2|_{-\infty }^{+\infty
  }/2$ which is zero for confined light beam and extended 2DES. Apparently,
  this trick does not apply to confined 2DES with non-zero field at the
  periphery.}\BibitemShut {Stop}%
\bibitem [{\citenamefont {Senior}\ and\ \citenamefont
  {Hartree}(1952)}]{Senior}%
  \BibitemOpen
  \bibfield  {author} {\bibinfo {author} {\bibfnamefont {T.~B.~A.}\
  \bibnamefont {Senior}}\ and\ \bibinfo {author} {\bibfnamefont {D.~R.}\
  \bibnamefont {Hartree}},\ }\bibfield  {title} {\bibinfo {title} {Diffraction
  by a semi-infinite metallic sheet},\ }\href
  {https://doi.org/10.1098/rspa.1952.0137} {\bibfield  {journal} {\bibinfo
  {journal} {Proc. Roy. Soc. London. A}\ }\textbf {\bibinfo {volume} {213}},\
  \bibinfo {pages} {436} (\bibinfo {year} {1952})}\BibitemShut {NoStop}%
\bibitem [{\citenamefont {Mikhailov}\ and\ \citenamefont
  {Ziegler}(2007)}]{Mikhailov_new_mode}%
  \BibitemOpen
  \bibfield  {author} {\bibinfo {author} {\bibfnamefont {S.~A.}\ \bibnamefont
  {Mikhailov}}\ and\ \bibinfo {author} {\bibfnamefont {K.}~\bibnamefont
  {Ziegler}},\ }\bibfield  {title} {\bibinfo {title} {New electromagnetic mode
  in graphene},\ }\href {https://doi.org/10.1103/PhysRevLett.99.016803}
  {\bibfield  {journal} {\bibinfo  {journal} {Phys. Rev. Lett.}\ }\textbf
  {\bibinfo {volume} {99}},\ \bibinfo {pages} {016803} (\bibinfo {year}
  {2007})}\BibitemShut {NoStop}%
\bibitem [{\citenamefont {Low}\ \emph {et~al.}(2012)\citenamefont {Low},
  \citenamefont {Perebeinos}, \citenamefont {Kim}, \citenamefont {Freitag},\
  and\ \citenamefont {Avouris}}]{Low2012}%
  \BibitemOpen
  \bibfield  {author} {\bibinfo {author} {\bibfnamefont {T.}~\bibnamefont
  {Low}}, \bibinfo {author} {\bibfnamefont {V.}~\bibnamefont {Perebeinos}},
  \bibinfo {author} {\bibfnamefont {R.}~\bibnamefont {Kim}}, \bibinfo {author}
  {\bibfnamefont {M.}~\bibnamefont {Freitag}},\ and\ \bibinfo {author}
  {\bibfnamefont {P.}~\bibnamefont {Avouris}},\ }\bibfield  {title} {\bibinfo
  {title} {{Cooling of photoexcited carriers in graphene by internal and
  substrate phonons}},\ }\href {https://doi.org/10.1103/PhysRevB.86.045413}
  {\bibfield  {journal} {\bibinfo  {journal} {Physical Review B}\ }\textbf
  {\bibinfo {volume} {86}},\ \bibinfo {pages} {045413} (\bibinfo {year}
  {2012})}\BibitemShut {NoStop}%
\bibitem [{\citenamefont {Xia}\ \emph {et~al.}(2009)\citenamefont {Xia},
  \citenamefont {Mueller}, \citenamefont {Golizadeh-Mojarad}, \citenamefont
  {Freitage}, \citenamefont {Lin}, \citenamefont {Tsang}, \citenamefont
  {Perebeinos},\ and\ \citenamefont {Avouris}}]{Xia2009a}%
  \BibitemOpen
  \bibfield  {author} {\bibinfo {author} {\bibfnamefont {F.}~\bibnamefont
  {Xia}}, \bibinfo {author} {\bibfnamefont {T.}~\bibnamefont {Mueller}},
  \bibinfo {author} {\bibfnamefont {R.}~\bibnamefont {Golizadeh-Mojarad}},
  \bibinfo {author} {\bibfnamefont {M.}~\bibnamefont {Freitage}}, \bibinfo
  {author} {\bibfnamefont {Y.~M.}\ \bibnamefont {Lin}}, \bibinfo {author}
  {\bibfnamefont {J.}~\bibnamefont {Tsang}}, \bibinfo {author} {\bibfnamefont
  {V.}~\bibnamefont {Perebeinos}},\ and\ \bibinfo {author} {\bibfnamefont
  {P.}~\bibnamefont {Avouris}},\ }\bibfield  {title} {\bibinfo {title}
  {{Photocurrent imaging and efficient photon detection in a graphene
  transistor}},\ }\href {https://doi.org/10.1021/nl8033812} {\bibfield
  {journal} {\bibinfo  {journal} {Nano Letters}\ }\textbf {\bibinfo {volume}
  {9}},\ \bibinfo {pages} {1039} (\bibinfo {year} {2009})}\BibitemShut
  {NoStop}%
\bibitem [{\citenamefont {Bandurin}\ \emph {et~al.}(2018)\citenamefont
  {Bandurin}, \citenamefont {Gayduchenko}, \citenamefont {Cao}, \citenamefont
  {Moskotin}, \citenamefont {Principi}, \citenamefont {Grigorieva},
  \citenamefont {Goltsman}, \citenamefont {Fedorov},\ and\ \citenamefont
  {Svintsov}}]{Bandurin2018}%
  \BibitemOpen
  \bibfield  {author} {\bibinfo {author} {\bibfnamefont {D.~A.}\ \bibnamefont
  {Bandurin}}, \bibinfo {author} {\bibfnamefont {I.}~\bibnamefont
  {Gayduchenko}}, \bibinfo {author} {\bibfnamefont {Y.}~\bibnamefont {Cao}},
  \bibinfo {author} {\bibfnamefont {M.}~\bibnamefont {Moskotin}}, \bibinfo
  {author} {\bibfnamefont {A.}~\bibnamefont {Principi}}, \bibinfo {author}
  {\bibfnamefont {I.~V.}\ \bibnamefont {Grigorieva}}, \bibinfo {author}
  {\bibfnamefont {G.}~\bibnamefont {Goltsman}}, \bibinfo {author}
  {\bibfnamefont {G.}~\bibnamefont {Fedorov}},\ and\ \bibinfo {author}
  {\bibfnamefont {D.}~\bibnamefont {Svintsov}},\ }\bibfield  {title} {\bibinfo
  {title} {{Dual origin of room temperature sub-terahertz photoresponse in
  graphene field effect transistors}},\ }\href
  {https://doi.org/10.1063/1.5018151} {\bibfield  {journal} {\bibinfo
  {journal} {Applied Physics Letters}\ }\textbf {\bibinfo {volume} {112}},\
  \bibinfo {pages} {141101} (\bibinfo {year} {2018})}\BibitemShut {NoStop}%
\bibitem [{\citenamefont {Gusikhin}\ \emph {et~al.}(2018)\citenamefont
  {Gusikhin}, \citenamefont {Muravev}, \citenamefont {Zagitova},\ and\
  \citenamefont {Kukushkin}}]{Gusikhin2018}%
  \BibitemOpen
  \bibfield  {author} {\bibinfo {author} {\bibfnamefont {P.~A.}\ \bibnamefont
  {Gusikhin}}, \bibinfo {author} {\bibfnamefont {V.~M.}\ \bibnamefont
  {Muravev}}, \bibinfo {author} {\bibfnamefont {A.~A.}\ \bibnamefont
  {Zagitova}},\ and\ \bibinfo {author} {\bibfnamefont {I.~V.}\ \bibnamefont
  {Kukushkin}},\ }\bibfield  {title} {\bibinfo {title} {{Drastic Reduction of
  Plasmon Damping in Two-Dimensional Electron Disks}},\ }\href
  {https://doi.org/10.1103/PhysRevLett.121.176804} {\bibfield  {journal}
  {\bibinfo  {journal} {Physical Review Letters}\ }\textbf {\bibinfo {volume}
  {121}},\ \bibinfo {pages} {176804} (\bibinfo {year} {2018})}\BibitemShut
  {NoStop}%
\end{thebibliography}%

\end{document}